\documentclass[onecolumn,12pt,trackchanges]{aastex62mtl}
\usepackage{graphics}
\usepackage{grffile}
\usepackage{rotating}
\usepackage{sidecap}
\usepackage{natbib}
\usepackage{url}

\bibpunct{[}{]}{,}{n}{}{;}

\usepackage{amsmath}  
\usepackage{amssymb,amsmath}  
\usepackage{color}
\usepackage{times}
\usepackage[normalem]{ulem}
\usepackage{enumitem}
\usepackage{cancel}
\usepackage{verbatim}
\usepackage{mathrsfs}
\usepackage[mathscr]{euscript}
\usepackage{fancyref}
\usepackage{hyperref}
\usepackage{physics}

\begin{document}

\setcounter{page}{0}

\title{{\LARGE
Astro2020 Project White Paper} \\
{
\begin{changemargin}{0cm}{0cm} 
\Large\bf The NANOGrav Program for Gravitational Waves and Fundamental Physics
\end{changemargin}
}
\hspace{-0.6in}\includegraphics[width=3in]{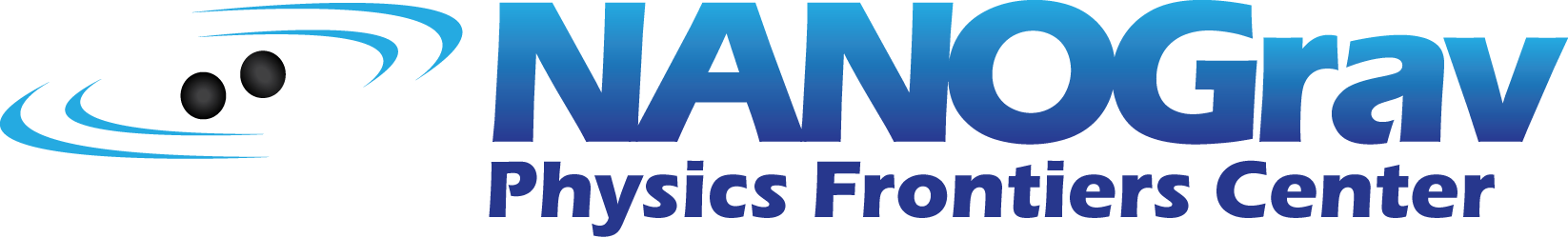}
~ \\ 
{\large The North American Nanohertz Observatory for  Gravitational Waves}
~ \\ ~ \\ 
{\rm 15 July 2019}
}

\shorttitle{NANOGrav}
\shortauthors{NANOGrav}

\section*{~~}
\vspace{-0.65in}

{\bf Thematic areas:} Multi-messenger astronomy and astrophysics; Cosmology and fundamental physics; Formation and evolution of compact objects.

{\bf Contact author:}
\vspace {-0.15in}
Scott Ransom 
(NANOGrav Chair), 
NRAO,
scott.ransom@nanograv.org
\vspace {0.2in}

    {\bf Authors:}
A.~Brazier (Cornell), S.~Chatterjee (Cornell), T.~Cohen (NMT), J.~M.~Cordes (Cornell), M.~E.~DeCesar (Lafayette), P.~B.~Demorest (NRAO),  J.~S.~Hazboun (UW Bothell), M.~T.~Lam (WVU, RIT), R.~S.~Lynch (GBO), M.~A.~McLaughlin (WVU), S.~M.~Ransom (NRAO), X.~Siemens (OSU, UWM), S.~R.~Taylor (Caltech/JPL, Vanderbilt), and S.~J.~Vigeland (UWM) for the NANOGrav Collaboration ($\sim 50$ institutions, 100$+$ individuals)

\begin{figure}[h]
    \centering

    \vspace{-3mm}    
    \includegraphics[width=5in]{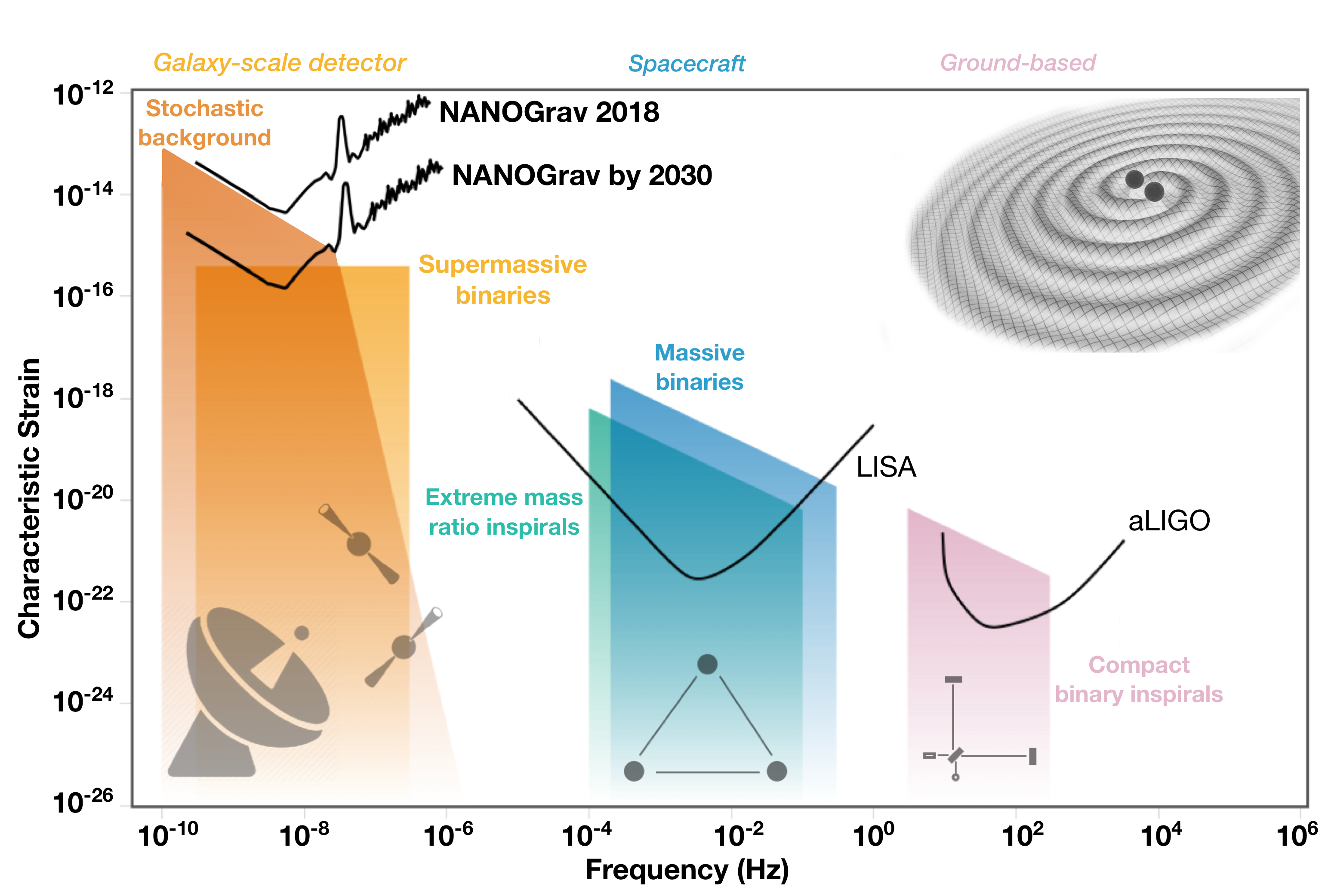}
    \vspace{-1mm}
    \label{fig:cover}
\end{figure}

{\bf Related NANOGrav Science White Papers}
\vspace{-3mm}
\begin{itemize}[leftmargin=*,label={$\cdot$}, itemsep=-3pt]
\item {\it Gravitational Waves, Extreme Astrophysics, 
and Fundamental Physics with Precision  Pulsar Timing}, J. Cordes et al.
\item {\it  Supermassive Black-hole Demographics \& Environments with Pulsar Timing Arrays},  S. Taylor et al.
\item {\it Fundamental Physics with Radio Millisecond  Pulsars}, E. Fonseca et al.
\item {\it Physics Beyond the Standard Model with Pulsar Timing Arrays}, X. Siemens et al.
\item {\it Multi-messenger Astrophysics with Pulsar Timing Arrays}, L.~Kelley et al.
\item {\it The Virtues of Time and Cadence for Pulsars and Fast Transients}, R. Lynch et al.
\item {\it Twelve Decades: Probing the ISM from kiloparsec to sub-AU scales}, D. Stinebring et al. 
\end{itemize}

\vspace{-6pt} {\footnotesize {\bf About the cover image:}
  Characteristic GW strain across the full spectrum. Binary black hole
  sources are shown in each frequency band, probed by current (PTA and
  ground-based) and planned (LISA) detectors. The allowed range of the
  stochastic background is scaled to be compatible with the NANOGrav
  2018 upper limit.  S.~Taylor \& C.~Mingarelli, adapted from
  gwplotter.org (Moore, Cole, Berry 2014) and based on a figure in
  Mingarelli \& Mingarelli (2018).  Illustration of merging black
  holes adapted from R. Hurt/Caltech-JPL/EPA.}

\clearpage

\section{Key Science Goals and Objectives}
\label{sec:obj}

NANOGrav (the North American Nanohertz Observatory for Gravitational Waves) exploits the high-precision timing of an array of Galactic millisecond pulsars (MSPs) in a bid to unveil timing deviations
induced by gravitational waves (GWs). A GW transiting an Earth--pulsar line of sight distorts the intervening spacetime, causing pulses to arrive before or after their expected times of arrival (TOAs). Crucially, such a GW will affect \textit{all} pulsars in the Galaxy, imprinting a correlated influence on the TOAs that allows NANOGrav and other pulsar timing arrays (PTAs) to effectively synthesize a kiloparsec-scale GW detector. The imprinted correlations are quadrupolar \cite{hd83}, reducing in magnitude as pulsars become more separated on the sky, then rising again at antipodal separation (see Figure~\ref{fig:HDseparations}, top panel). No other known process produces these ``Hellings \& Downs'' correlations; clock-standard drifts produce the same errors in all pulsars \citep{2012MNRAS.427.2780H}, while TOA-barycentering errors produce dipolar correlations \citep{2010ApJ...720L.201C}. 

NANOGrav now has $T \sim 15$ years of high-precision data from many pulsars, with new ones being regularly added to give a current total of 77 \cite{11yrdataset}. Each pulsar is observed with the Green Bank Telescope (GBT) or Arecibo every $\Delta t \sim 1-3$ weeks with an integration time of $\sim 20$ minutes to produce TOAs. We are therefore sensitive to GW frequencies in the range of $\sim$ 2~nHz to 1~$\mu$Hz (i.e. $1/T < f < 1/2\Delta t$). The dominant source of GWs in this range is a cosmological population of supermassive binary black holes (SMBBHs; \cite{phinney2001,jaffebacker,sesana13}) that form as a by-product of hierarchical galaxy growth, where galaxies (each with central massive black holes \cite{kh13,mm13}) merge repeatedly over cosmic time to produce ever larger galaxies.
As such, the mergers of massive galaxies trigger the slow pairing and merger of the massive black holes at their cores. Below, we list our key scientific objectives, followed by further discussion of these items. 

\textbf{NANOGrav's key objectives are to:}
\begin{enumerate}[noitemsep,topsep=0pt,label={(\bfseries O\arabic*)}]
    \addtolength{\itemindent}{-5pt}
    \item Detect a stochastic background of nanohertz gravitational waves;
    \item Characterize this background in terms of the demographics of the expected emitting population of SMBBHs;
    \item Detect individual SMBBHs, and perform multi-messenger follow-up with large synoptic photometric and spectroscopic surveys operational in the 2020s+;
    \item Constrain fundamental physics by probing the neutron star  equation of state, cosmic strings, primordial GWs, first-order cosmological phase transitions, beyond-GR theories of gravity, and dark matter.
\end{enumerate}

\begin{figure}[!h]
    \centering
    \includegraphics[width=\textwidth]{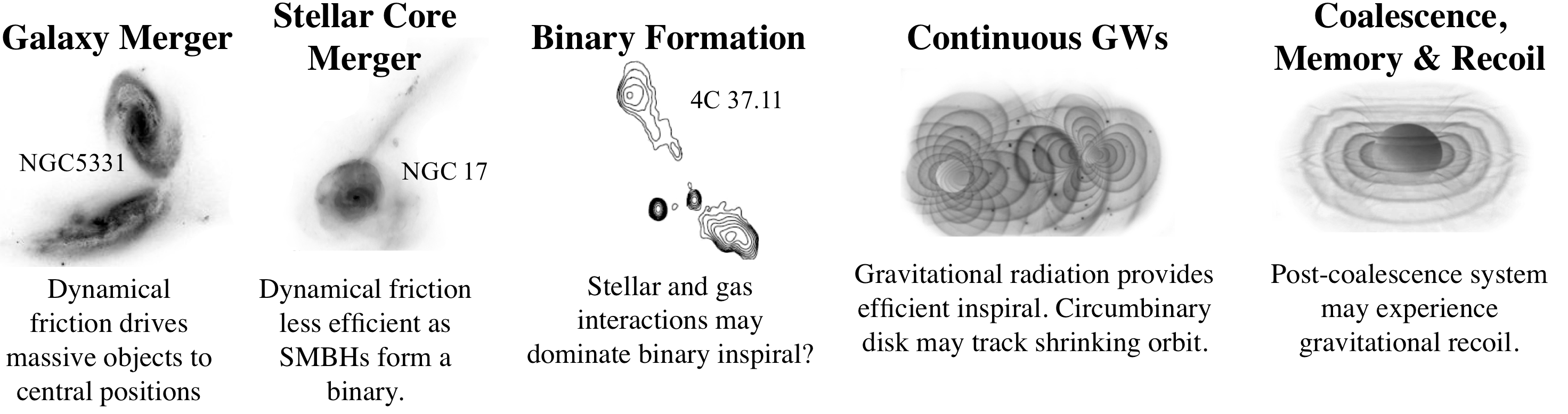}
    \caption{The evolution of SMBBH mergers through galaxy interactions. See discussion in Section~\ref{sec:obj}. This has been modified from Figure 3 in Burke-Spolaor et al. 2019~\cite{SBS2019}. Image  credits: Hubble/STSci;  Rodriguez et al. 2006~ \cite{Rodriguez2006};   C. Henze/NASA; C. Cuadra.  
    }
    \label{fig:galevol}
\end{figure}

\begin{figure}
    \centering
    \includegraphics[width=0.49\textwidth,height=150pt]{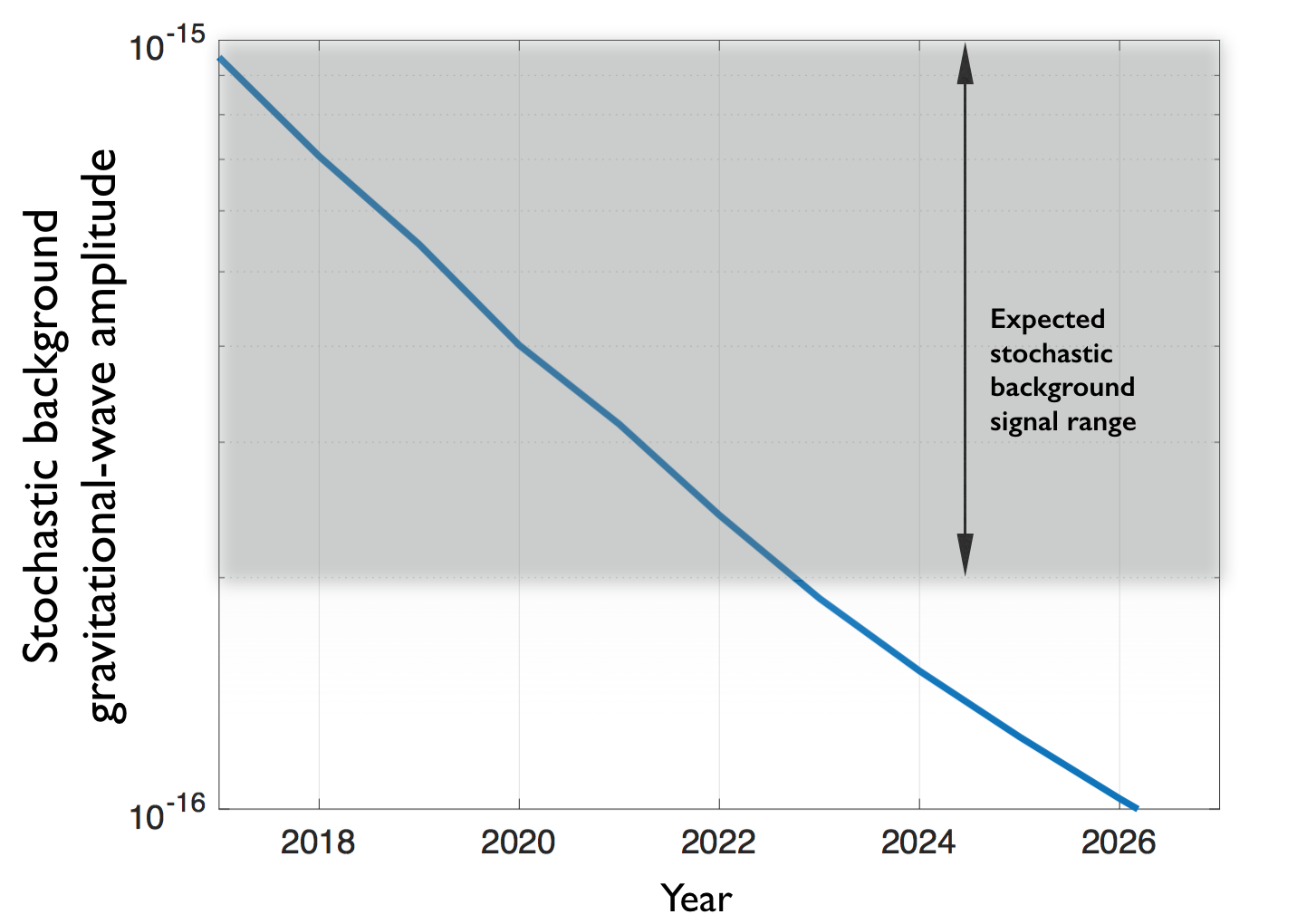}
    \includegraphics[width=0.5\textwidth]{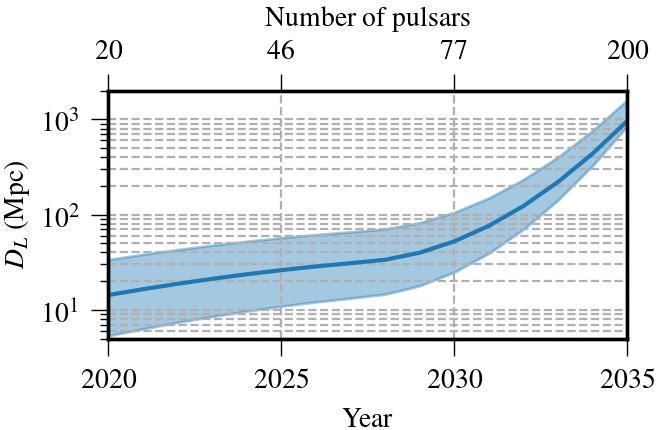}
    \caption{ \textit{Left:} Detection projection for the nanohertz background of GWs from SMBBHs for 50\% of realizations of the Universe. After detection, efforts will move towards characterization of the background, providing unique insights into galaxy growth and evolution. \textit{Right:} Projected sensitivity to GWs from individual SMBBHs with $f_\mathrm{gw}$ = 3 nHz and $\mathcal{M} = 10^9 \, M_\odot$. The band denotes the distance out to which an individual source can be detected, depending on the source's sky location and the distribution of pulsars; the solid line shows the sky-averaged distance. Here, the number of pulsars only includes those with more than $T=10$ years of observations, since MSPs are only sensitive to GWs with $f_\mathrm{gw} \geq 1/T$. See discussion in Section~\ref{sec:obj}.}

    \vspace{3mm}

    \label{fig:cw_sensitivity}
\end{figure}

\textbf{O1}: \textbf{\textit{The first PTA detection will likely be a stochastic background of GWs}} emanating from the entire population of inspiraling SMBBHs. Figure~\ref{fig:galevol} shows the SMBBH merger timeline, with PTAs sensitive to the early inspiral and merger memory burst of SMBBHs. Under fiducial assumptions of a circular GW-driven binary population, the characteristic strain spectrum of this stochastic GW signal follows $h_c(f)\propto f^{-2/3}$ (e.g.~\citep{phinney2001}). The resulting timing deviations, $\delta t$, are therefore predominantly low-frequency, with a power spectrum $S_{\delta t}(f)\propto f^{-13/3}$ yr$^3$. We have developed a sophisticated analysis pipeline that models the correlated influence of this GW background (GWB) both temporally (across years of observations) and spatially (between pulsars) to assess the evidence of this signal in our datasets, and to interpret constraints in the absence thereof \cite{nano11yr_gwb}. Based on rigorous forecasting of pulsar noise properties, PTA expansion through future surveys, and SMBBH population properties, we expect that NANOGrav will make a robust detection (Figure~\ref{fig:cw_sensitivity}, left) of the stochastic GWB within the next $3-7$ years \cite{sej+13,rsg15,taylor16,2017MNRAS.471.4508K}. 

\textbf{O2}: \textbf{\textit{In the immediate pre-detection and post-detection era, NANOGrav will constrain, then measure, the spectral characteristics of this GWB}}. While fiducial assumptions give the power-law strain signal mentioned above, the true spectrum will likely turn over at lower frequencies (i.e. larger orbital separations) due to SMBBHs remaining dynamically coupled to their ambient galactic environments (e.g. binary--star scattering events, circumbinary disk coupling) (e.g.~\citep{2013CQGra..30v4014S,SBS2019}).
The slope and frequency of this spectral turnover maps to the typical astrophysical conditions in galaxy cores (e.g.~\citep{2015PhRvD..91h4055S,2017PhRvL.118r1102T,2019MNRAS.tmp.1679C}). Characterizing this low-frequency spectral behavior will allow NANOGrav to probe the final-parsec dynamical interactions of SMBBHs with their galactic environments, while the overall signal amplitude will arbitrate rate factors such as BH--host-galaxy scaling relationships \citep{2016ApJ...826...11S}, dynamical friction timescales, BH occupation fractions, etc. (see NANOGrav Science White Paper by Taylor et al.~\cite{WP_Taylor}).

\textbf{O3}: \textbf{\textit{NANOGrav will individually resolve sufficiently nearby and/or massive SMBBHs against the backdrop of the full population}}, allowing us to localize the binary and characterize its orbital geometry. We expect to detect one or more individual SMBBH as a quasi-monochromatic GW signal by 2030 \cite{rsg15,mls+17,kbh+18}, with redshift $z\lesssim 0.5$, chirp mass $10^9~M_\odot < \mathcal{M} < 10^{10}~M_\odot$ \citep{rsg15}, and sky localization $\Delta\Omega \sim 10^2-10^3$ deg$^2$ \citep{2016ApJ...817...70T,2019MNRAS.485..248G} (Figure~\ref{fig:cw_sensitivity}, right). Individual SMBBH GW searches will be cross-validated with near-future EM surveys that can register photometric (e.g. LSST) and spectroscopic (e.g. SDSS-V) signatures of binary AGN (see e.g.~\citep{2019MNRAS.485.1579K} and references therein), allowing for a complete multi-messenger portrait of merging supermassive black holes (see NANOGrav Science White Paper by Kelley et al.~\cite{WP_Kelley}). Indeed, while PTAs are insensitive to the high-frequency oscillatory component of the final SMBBH merger, GR predicts a constant metric offset caused at the time of the merger itself that is broadband and potentially detectable by PTAs (e.g.~\citep{2019arXiv190611936I,2010MNRAS.401.2372V,2012ApJ...752...54C,2015ApJ...810..150A}).

\textbf{O4}: \textbf{\textit{Beyond binary black holes, PTAs such as NANOGrav are also sensitive to a variety of early-Universe GW sources}} (see e.g.~\citep{WP_Siemens} and references therein), such as a primordial GW background from inflation, cosmic string networks formed during phase transitions, and collisions of vacuum bubbles formed during a cosmological first-order phase transition (e.g. the QCD transition). Additionally, the many Earth-pulsar lines-of-sight constituting a PTA offer a way to test modified theories of gravity through GW polarization states beyond the usual $+$ and $\times$ (e.g.~\citep{2008ApJ...685.1304L,2012PhRvD..85h2001C,2015PhRvD..92j2003G,2018PhRvL.120r1101C}). Likewise, \textbf{\textit{PTAs generate a high output of synergistic science that is not directly related to GWs}}, including studies of the extreme physics of neutron stars and the nuclear equation of state (see NANOGrav Science White Paper by Fonseca et al.~\cite{WP_Fonseca}), characterization of the ionized interstellar medium (see NANOGrav Science White Paper by Stinebring et al.~\cite{WP_Stinebring}), constraints on cold dark matter sub-structure (e.g.~\citep{2011PhRvD..84d3511B,2019arXiv190104490D}), and even constraints on an ultra-light scalar field component of the Milky Way dark matter halo \citep{2018PhRvD..98j2002P} (see NANOGrav Science White Paper by Siemens et al.~\cite{WP_Siemens}).

\clearpage
\section{Technical Overview}

Precision pulsar timing involves modeling observed arrival times over time spans of years and seeking departures from predicted arrival times to measure orbital motions (including general relativistic corrections), determine masses of neutron stars and their companions and, of course, to detect GWs.  
The successes of pulsar timing are built on an empirical foundation of pulsar phenomenology: 
\begin{enumerate}[topsep=0pt,itemsep=-2pt,partopsep=-12pt]
\itemsep -2pt
\item
The high rotational stability of pulsars allows prediction of individual pulse arrival times to $\lesssim$~microsecond precision months in advance, such that low duty-cycle monitoring (weekly to monthly) is sufficient;
\item
Emitted pulses have phases locked to the spin of the neutron star, on average;
\item
Average pulse shapes   have stable, predictable  radio frequency dependences;
\item
Propagation through the interstellar medium (ISM) is largely regular and predictable according to the cold plasma dispersion law.
\end{enumerate}
 NANOGrav (and other PTA groups around the world) use MSPs for GW detection because they are the most spin-stable neutron stars, and because their few-millisecond spin periods allow the most pulses to be averaged in a telescope hour; this pulse-averaging increases the signal-to-noise ratio (S/N) and reduces pulse phase jitter intrinsic to the emission process.  We choose 
 MSPs with the best timing precision: those that are bright, have the narrowest pulse features, and have minimal noise in their spin rates.  Best practices in pulsar timing exploit all of these features to provide RMS timing precision better than 100~ns in some cases.

\subsection{Pulsar Timing}
\label{sec:pulsar_timing}

NANOGrav's science program  has the following requirements for GW detection with precision pulsar timing (see  Science White Paper by Cordes \& McLaughlin~\cite{WP_Cordes}):

\textbf{\textit {A large number of MSPs}}  to average out astrophysical noise peculiar to each pulsar (spin noise and pulse-phase jitter) and each line of sight (interstellar variance from  stochastic dispersion and scattering).
NANOGrav's current 77-MSP program provides important constraints on SMBBH inspiral astrophysics and important results in synergistic science.    The goal is to time 200 MSPs by the mid-2020s to increase sensitivity to all nanohertz GW signal types; modeling of the Galactic MSP population predicts this number of high-quality MSPs will likely be accessible to future telescopes.
Our ongoing and new pulsar surveys, planned instrumentation upgrades, and new partnerships will yield this increase.

\textbf{\textit{A uniform sky distribution of MSPs}}  to sample the quadrupolar response to GWs and to allow detection of anisotropies in the stochastic GW background \cite{msmv2013,tmg+2015}. 
NANOGrav's current sky distribution of MSPs is highly non-uniform (see current status in Section~\ref{sec:org}), with concentrations and voids in the North and a complete lack of coverage in the far South.  Pulsar surveys (including ongoing searches in currently low-MSP-density sky regions) and new partnerships, from now through the mid-2020s, aim to fill these voids, leading to more robust sensitivity to the quadrupolar signature.

\textbf{\textit{Large collecting area telescopes:}} Pulsar timing requires the largest available (and planned) $\sim$GHz-frequency radio collecting areas to provide the highest signal-to-noise and thus smallest time-of-arrival errors. 
\textbf{\textit{Arecibo and the GBT are currently the best telescopes for PTA work; it is crucial that NANOGrav maintain access to both through the 2020s.}}
The FAST telescope in China will contribute on some objects but is unlikely to provide sufficient cadence on enough pulsars.    The planned Deep Synoptic Array with 2000 elements (DSA-2000) is designed to be an excellent pulsar-timing instrument as well as a radio transients instrument.  The proposed Next Generation Very Large Array (ngVLA) will have pulsar timing capabilities and provide complementary coverage of higher radio frequencies.

\textbf{\textit{A minimum integration time per epoch}}  to average $10^5$ -- $10^6$ single pulses needed for convergence to
the template pulse profile used in matched filtering estimation of arrival times.
Typically about one hour per epoch per MSP is  split between two observing bands.    Integration times will be doubled  by the imminent development of ``ultra-wideband'' (UWB) feed-antenna/receiver systems for the GBT and Arecibo.    Telescopes that may come online in the 2020s will also provide wideband capabilities. 

\textbf{\textit{Observing cadence of tens of epochs per year}} for adequately modeling astrometric  parameters and tracking time-variable interstellar delays (see NANOGrav Science White Papers by Lynch et al.~\cite{WP_Lynch} and Stinebring et al.~\cite{WP_Stinebring}).
The current cadence is three to four weeks on most MSPs and one week on a few objects to increase CW sensitivity and allow better mitigation of interstellar effects.   The goal is  a one-week cadence on most MSPs by the second half of the 2020s. 

\textbf{\textit{Timing continuity}} is crucial to GW detection because  GW signals and some types of TOA noise are time-correlated.
Gaps in the timing program longer than about a month must be avoided, either in the availability of  a telescope or for a particular instrument.    New instruments on a telescope need to be used in parallel with previous generation instruments to calibrate  timing offsets at the sub-100~ns level. 

\textbf{\textit{Broad radio frequency (RF) coverage ($\sim$0.4 to 3 GHz)}} to disentangle and mitigate interstellar delays and intrinsic RF dependence of pulse shapes, and to increase pulse signal-to-noise ratios. 
Most current receiver systems are limited to octave bandwidths (e.g., 2:1 frequency ratio) so
two RF bands are currently used  ($\sim$100-MHz bandwidth for RF $<$~1 GHz and 
$\sim$800-MHz bandwidth at higher frequencies) in separate observations.    New UWB feeds under development at Arecibo and the GBT covering $\sim$0.7--3~GHz will improve the instantaneous bandwidth and require only a single observation per epoch.   The recently-completed CHIME telescope complements these with lower-frequency coverage (0.4--0.8~GHz).  Future telescopes (e.g. DSA-2000 and the ngVLA) also will provide UWB TOAs. Some of the new MSPs will require timing at high frequencies  to combat larger interstellar noise expected from their higher dispersion measures (i.e.~at large distances or deeper in the Galactic plane).

\textbf{\textit{Mitigation of RF interference (RFI):}} Narrowband, impulsive, and swept frequency RFI is excised as part of the NANOGrav processing pipeline.   However, the numbers of transmitters and RFI signal types are growing. 
NANOGrav works with Arecibo and GBT staff to mitigate RFI in receiver systems and negotiate with transmitting entities, where possible,  for radio quiet times.   RFI mitigation in the data pipeline requires continuous attention and appropriate code development. 

\textbf{\textit{Accurate solar-system ephemeris}} to reference arrival times to the solar system barycenter.    
Current JPL ephemerides are accurate to about 50~ns over a ten-year period and show differences related to Jupiter's orbital period. Additional spacecraft flybys of the gaseous planets combined with longer pulsar timing data spans  will improve the ephemeris.

Detection of any type of nanohertz GW signal is improved with increases in telescope sensitivity, integration time per epoch, frequency coverage, and cadence. Different  GW signals emphasize the importance of  specific aspects of the above requirements.   Detection of the stochastic background, for example,  is strongly dependent on the number of MSPs and their sky distribution \citep{sej+13} while  single- source detections (CW and burst signals) also favor the lowest RMS errors in arrival times \citep{blf11,cal+14}.  

\subsection{GW Detection Methodology}
\label{sec:detection}

\begin{figure}[ht!]
    \centering
    \begin{minipage}[c]{0.54\textwidth}
    \includegraphics[width=\textwidth]{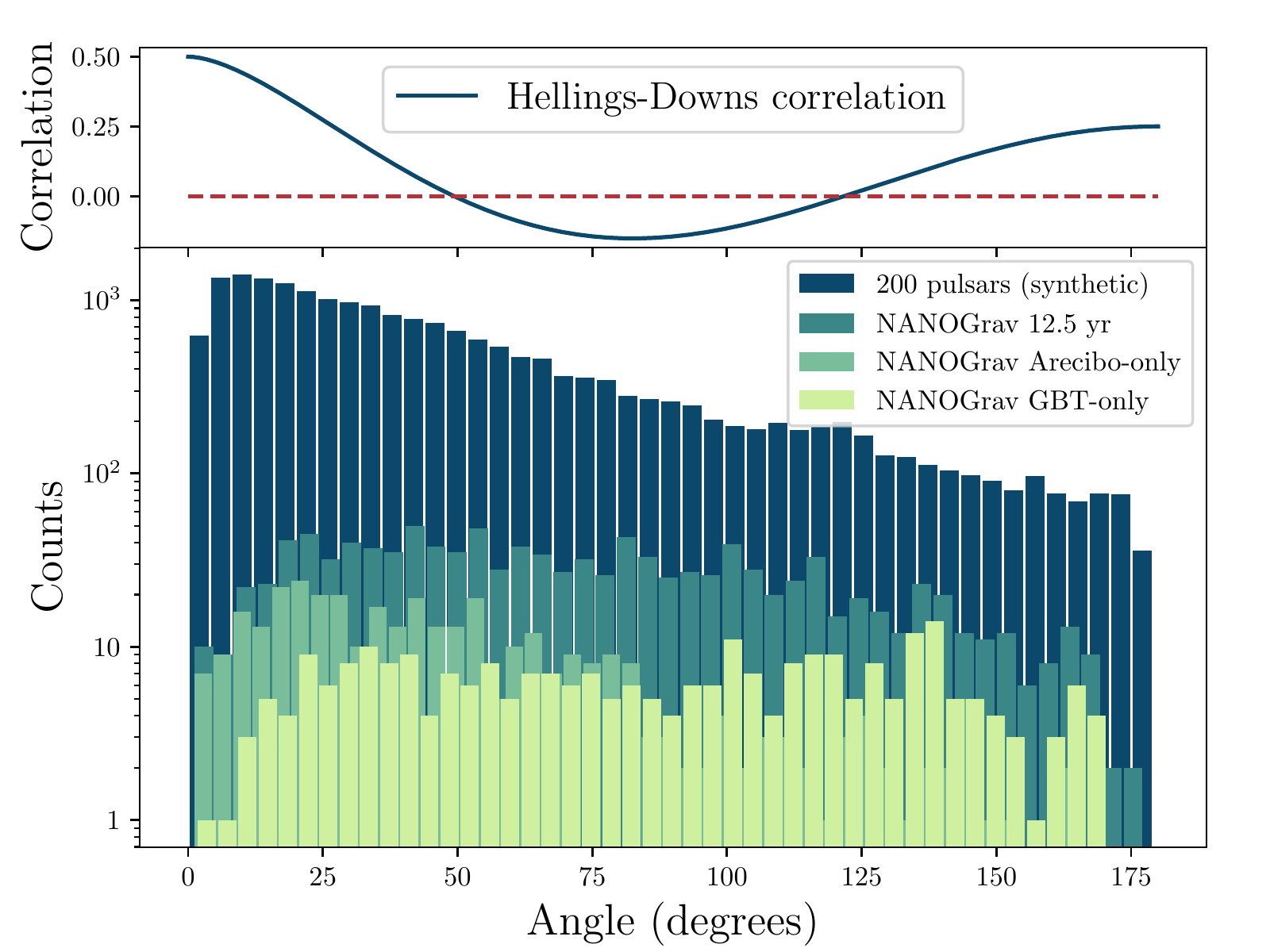}
  \end{minipage}\hfill
  \begin{minipage}[c]{0.45\textwidth}
    \caption{The Hellings and Downs correlation function and its sampling with pulsar pairs by the NANOGrav PTA.  The upper panel shows the H-D curve in solid blue and zero correlation in dashed red. The lower panel shows the number of pulsar pairs per angular separation bin for the 12.5-year NANOGrav data (GBT only, Arecibo only, both telescopes), as well as the sampling for a projected 200-pulsar PTA based on our MSP population simulations (T. Cohen, P. Demorest). Adding new pulsars to expand the array to 200 timed pulsars produces {\em over an order of magnitude} improvement in the number of samples per angular separation bin. See discussion in Sections \ref{sec:pulsar_timing} and \ref{sec:tech}.}
  \label{fig:HDseparations}
  \end{minipage}

\end{figure}

The key to detecting GWs with PTAs is to calculate residuals, the differences between measured and model TOAs, and identify  a signal that is common across many different pulsars and consistent with the properties of GWs.
To do so involves careful modeling of deterministic contributions to arrival times (spin, orbital, astrometric, and propagation terms) and of stochastic astrophysical and instrumental delays, and removal of systematic errors.  Stochastic noise includes white noise contributions from pulse jitter, interstellar scintillation, and measurement errors while steep-spectrum red noise  ($f^{-2}$ to $f^{-6}$) is from neutron star spin fluctuations and from propagation through interstellar turbulence. 
In NANOGrav's GW analysis pipeline, we model the residuals of each pulsar with white and red noise combined with  the GW signal. The GWB signal is a red-noise process  that differs spectrally from astrophysical red noise and is common to all of the pulsars.   A powerful tool for identifying GWs is the two-point cross correlation of the residuals  between pulsar pairs, the Hellings-Downs (HD) correlation \cite{hd83}.    Figure \ref{fig:HDseparations} shows the HD curve (top) and the number of pulsar pairs vs. angular separation for NANOGrav's current PTA and the proposed PTA for the 2020s (bottom).  The HD curve reflects the  quadrupolar nature of GWs that differs from systematic clock  and ephemeris errors, which  also affect multiple pulsars but have monopolar and dipolar spatial correlations, respectively \citep{thk+16}.

A similar approach is taken for GWs from single sources that can be detected either as continuous wave (CW) signals or as bursts.   Detection of all three kinds of GWs  benefits from longer PTA data spans through standard noise averaging.  However,    `memory' bursts and the GWB particularly benefit because their signal strengths  grow with data span length  as $T$ and $T^{5/3}$, respectively.

We use a combination of Bayesian and frequentist detection methods to search for both stochastic and deterministic signals. For the GWB, we perform Bayesian searches for common red noise with and without quadrupolar spatial correlations and compare the Bayes factors. We also use  a frequentist estimator for the strength of the GWB that fits the cross-correlation to the Hellings-Downs curve \citep{abc+2009, dfg+2013, ccs+2015}. 
To address various issues, we  have also developed a hybrid Bayesian-frequentist detection statistic
\citep{vit+2018}.

\begin{figure}[b]
\centering
\vspace{-5pt}
\begin{minipage}[c]{0.55\textwidth}
    \includegraphics[width=\textwidth]{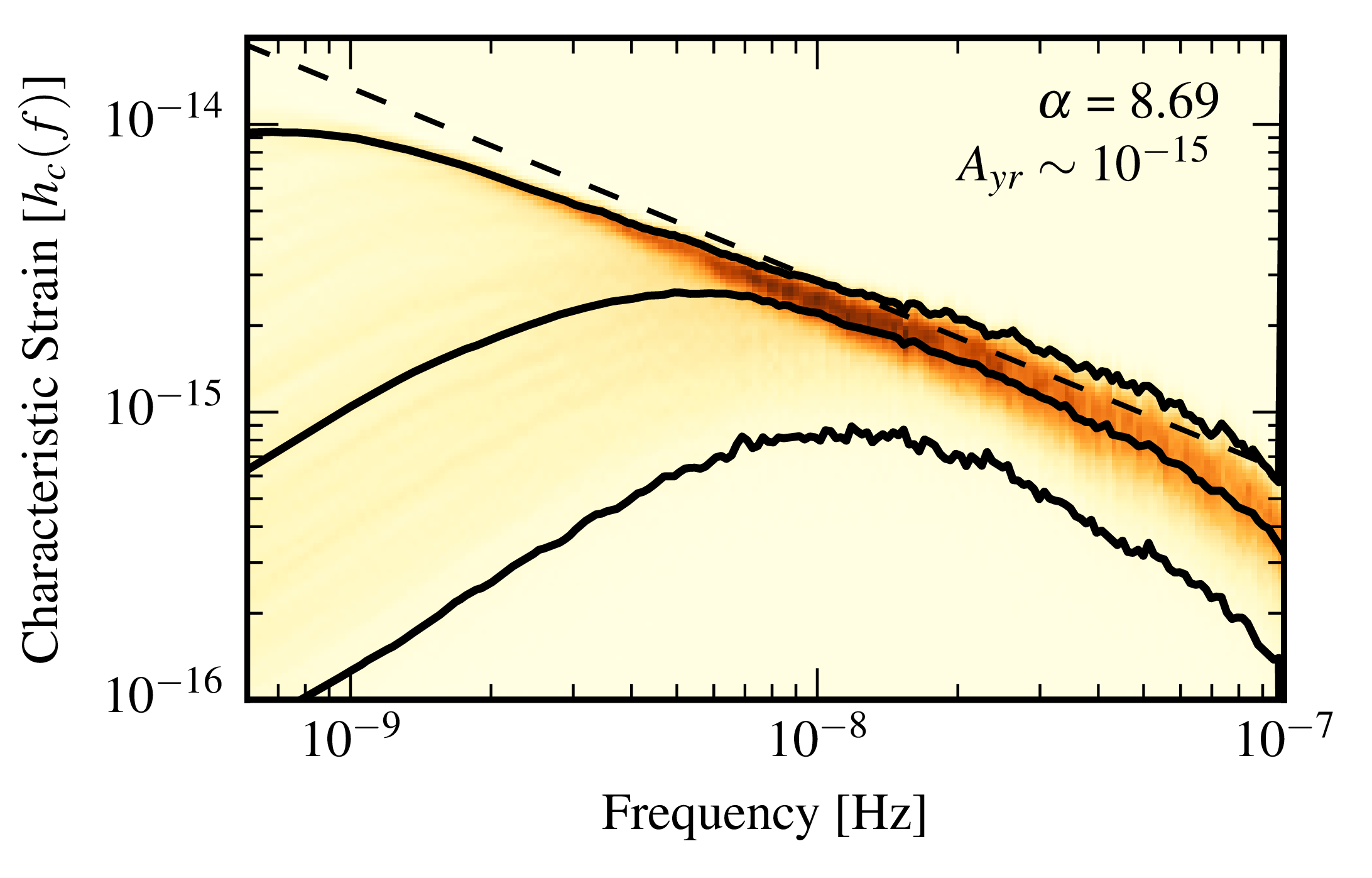}

\end{minipage} 
\hfill
\begin{minipage}[c]{0.43\textwidth}
   \caption{Upper limits on the GW stochastic background from SMBBH at 95\% confidence, as derived from the NANOGrav 11-yr data set. Results are shown assuming the power-law spectrum characteristic of SMBBHs  (straight dashed line). At each frequency we show the probability density of the strain, where the solid lines mark the 2.5\%, 50\%, and 97.5\% confidence levels, and assuming a specific value of $\alpha$, the y-intercept for the black-hole-to-bulge-mass relation in galaxies \cite{kh13}. See discussion in Section~\ref{sec:detection}. Figure adapted from NANOGrav 2019 \cite{nano11yr_gwb}.\label{fig:11yr_results}}
\end{minipage}
\vspace{-10pt}
\end{figure}

Our most recent constraints on the GWB come from our 11-year data set \cite{nano11yr_gwb}. Figure~\ref{fig:11yr_results} shows that our upper limit is now constraining models for the GWB involving the cosmic population of SMBBHs.   The same data set
has been used to constrain CW \cite{nano11yr_cw} and burst signals (Aggarwal et al., in prep). In the future our goal
is to  exploit   both the Earth terms and the pulsar terms that arise from GWs passing both ends of each line of sight.  This requires pulsar distance determinations to within a GW wavelength ($\lambda_\mathrm{gw} = 0.1 - 10 \, \mathrm{pc}$) to allow correction  for the light travel time between the pulsars and the Earth. More precise pulsar distances deriving from parallax measurements through timing or radio interferometry are needed to achieve this accuracy \citep{jkc+2018, mab+2018, dgb+2019}.

We expect to detect the GWB within the next several years \citep{tve+16} (see Section~\ref{sec:schedule} for our planned science schedule). By characterizing the properties of the GWB, we will study the population of SMBBHs which make up the background and the properties of their host galaxies. We have already placed limits on astrophysical models using our upper limits on the GWB. We expect to detect {\it individual} SMBBHs in the next 10 to 15 years \citep{rsg15,mls+17,kbh+18}. With 200 MSPs in the array, we expect to be sensitive to individual sources with chirp masses of $10^9 \, M_\odot$ out to 800 -- 1600 Mpc, depending on the sky location of the source and positions of the MSPs (see Figure~\ref{fig:cw_sensitivity}, right panel).

\textbf{\textit{Noise modeling:}} NANOGrav and our IPTA partners have developed sophisticated noise models as input to GW detection pipelines \citep[][]{2013CQGra..30v4002C, 2016ApJ...819..155L,2017ApJ...834...35L,  2016ApJ...821...66L, 2016MNRAS.457.4421C,  2016MNRAS.458.2161L}.  These take into account pulsar-intrinsic effects, including the motions of emission regions in pulsar magnetospheres that produce pulse jitter and subtle spin variations from variable torques acting on the neutron star.   They also include several processes from interstellar (also interplanetary and ionospheric) propagation, that produce variable delays from changes in dispersion and scattering with epoch.   Arrival times are obtained through standard matched filtering using time-averaged pulse profiles as templates, yielding errors due to finite signal-to-noise ratios and from small changes in pulse shapes from pulse jitter and interstellar scattering.   Finally, the primary systematic error due to the solar system ephemeris, discussed earlier,  will improve significantly with time.

PTA noise modeling techniques need continual development. Mitigation of noise sources requires characterization of their physical mechanisms, such as discrete small-scale structures in the ISM causing deterministic refraction events or magnetospheric/superfluid dynamics causing neutron star spin variations. Noise modeling also provides important input for inclusion of new MSPs into the PTA, which is an important part of our plan for the 2020s, in which we wish to select MSPs with narrow pulses, well-modeled lines of sight, and low levels of spin noise.

\section{Technology Drivers}
\label{sec:tech}

Three primary technical drivers for the NANOGrav program next decade are discussed in the following.

\textbf{\textit{Growing the PTA:}}
Realizing the GW sensitivity needed to meet NANOGrav's scientific goals requires adding new pulsars to the PTA.  This can be accomplished in three ways:
\begin{itemize}[topsep=0pt,itemsep=-2pt,partopsep=-12pt]
    \item{Conducting sensitive surveys for new high-precision MSPs;}
    \item{Timing known but faint MSPs with more sensitive telescopes in the future to make them viable;}
    \item{Incorporating data from Southern hemisphere telescopes to increase declination coverage.}
\end{itemize}

Practical wide-area surveys require the high  survey speeds provided by phased array feeds (PAFs) operating at $\sim$1.5 GHz.  The Focal L-Band Array for the GBT (FLAG) demonstrates the ability of PAFs to reach the required sensitivity.  Building on the success of FLAG, the Arecibo L-band Phased-Array Camera (ALPACA) PAF and beamformer for Arecibo will be the successor to the ALFA 7-beam receiver, enabling the next generation of sensitive Arecibo surveys.  A successor to FLAG that offers a wider field of view will provide crucial sky-coverage outside the declination range of Arecibo, FAST (China), and Southern hemisphere telescopes.  Deep follow-up of candidate pulsars detected in other EM bands and imaging surveys is complementary to wide-area surveys, e.g. radio observations of unidentified \textit{Fermi} sources.  Regardless of the approach, it is crucial that we have access to sensitive radio telescopes with sufficient time to carry out these surveys.

The sensitivity of DSA-2000, ngVLA, and international telescopes such as FAST and SKA will make pulsars that are currently S/N limited viable for inclusion in PTAs as soon as these telescopes come online.  MeerKAT (South Africa) and SKA in particular will also provide access to pulsars in the far South through NANOGrav's membership in the IPTA.  Full-sky coverage will be especially important for single-source detection and follow-up.

\textbf{\textit{Ultra-wideband Instrumentation:}}
As described in Section~\ref{sec:pulsar_timing}, wide-bandwidth observations are important for characterizing chromatic interstellar propagation delays and pulse-shape evolution and for increasing S/N.  The next generation of UWB radio receivers installed on the GBT and Arecibo will offer several advantages over the current strategy of using two separate receivers:
\begin{itemize}[topsep=0pt,itemsep=-2pt,partopsep=-12pt]
    \item{More precise measurement of propagation delays at each observing epoch;}
    \item{A factor of two improvement in observing efficiency by using a single radio receiver;}
    \item{Better S/N achieved through wider instantaneous bandwidth.}
\end{itemize}
The combined effect of these benefits can lead to a 2x improvement in TOA precision.  GBO is developing a 0.7--4 GHz receiver for the GBT (partially funded by the Gordon and Betty Moore Foundation), and a similar receiver is planned for Arecibo.  The basic design and associated technologies (amplifiers, cryogenics, etc.) have been successfully demonstrated in a UWB receiver deployed on the Parkes Telescope \cite{dbb+15}.  A straightforward expansion of existing digital spectrometers will provide the necessary observing modes for precision pulsar timing.

UWB receivers for GBT and Arecibo will be cryogenically cooled to achieve the best possible system noise.  An important technology driver for DSA-2000 are broadband low-noise amplifiers that operate at non-cryogenic temperatures.  These have the potential to achieve a total system noise competitive with cryogenic receivers while greatly reducing cost and complexity.

\begin{figure}[t!]
\centering
\begin{minipage}[c]{0.52\textwidth}
    \includegraphics[width=\textwidth]{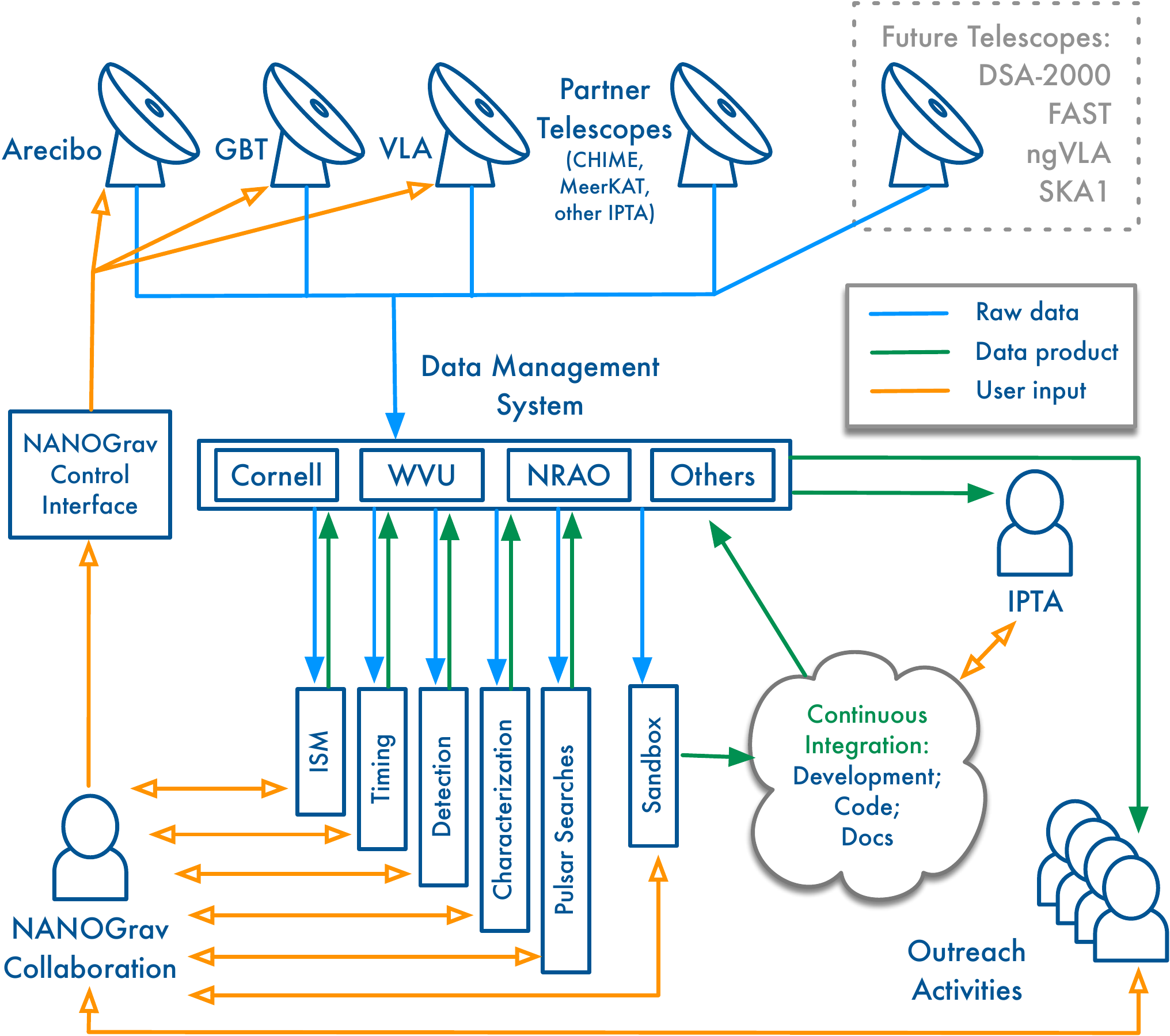}
\end{minipage} 
\hfill
\begin{minipage}[c]{0.45\textwidth}
    \caption{Data management and curation for the NANOGrav PTA showing the flow of data from telescopes to working groups within NANOGrav and to partners within the IPTA. Our framework can scale to much larger data volumes, especially for those predicted for future telescopes (e.g. DSA-2000, ngVLA, and international partner telescopes). Data archiving and curation occurs at two geographically distinct archives, with automated synchronization and checking of data integrity, and access to one of the archives---the master copy---is limited. This also serves our data and pipelines to the broader astronomical community. See discussion in Section~\ref{sec:tech}.
    }
    \label{fig:cyber}
\end{minipage} 
\end{figure}

\textbf{\textit{Data Curation and Cyber-infrastructure:}}
As a long-term, data-dependent project, NANOGrav data and data products are securely archived and curated indefinitely, with access provided to different communities including not just those interested in GW detection and multi-messenger astrophysics (see Science White Paper by Chang et al.~\cite{WP_Chang}) but also those interested in pulsar/neutron-star astronomy, ISM physics, and other scientific domains which can be probed with high-precision pulsar timing. Access to the data is limited for proprietary data and made available to the wider community at proprietary period expiration.

NANOGrav currently possesses an operational cyberinfrastructure framework, shown in Figure~\ref{fig:cyber}. Data flow directly from our telescopes and automatically populate our mirrored data archives. We automatically produce preliminary data products and we require that all processing from calibration of raw data to science-quality analyses must be reproducible, from calibration of raw data to the gravitational wave detection analysis. In particular, it is possible to re-run the entire pipeline on the raw data to produce the same results, and our pipeline will be distributed with all future releases of results. We also require simple creation of the correct processing environment, in our case encapsulated in a transferable format such as a software container.

The study, design, and implementation of new computational methods, including machine learning methods for pulsar survey analysis, is an important component of NANOGrav's ongoing development. Cyberinfrastructure support is essential for integrating new developments and technical approaches into the NANOGrav pipelines through testing, continuous integration, and tracking the provenance of the resulting data products.

NANOGrav's outreach includes workshops and demonstrations to illustrate NANOGrav science in its computational context, such as through Jupyter Notebooks. These notebooks  access  NANOGrav data in its entirety, subject to user permissions, including requesting data matching particular criteria. Management of the outreach computational platform will scale to using cloud resources for times of heavy demand from students and scientists.

\section{Organization, Partnerships, and Current Status}
\label{sec:org}

\textbf{\textit{Organization:}} NANOGrav is a mature and growing scientific collaboration with stable organizational structures and policies and effective shared management responsibilities (see the various sections of our website at \url{nanograv.org}).  NANOGrav's day-to-day operations and research activities are accomplished via eight different working groups, and coordination and management of the collaboration is the responsibility of an elected Management Team, which includes the NANOGrav Chair and the Co-Directors of the  NANOGrav  Physics Frontiers Center (PFC).  The NANOGrav PFC, currently in its fifth year, is a distributed center funded primarily through NSF Physics, with additional funding from NSF   Astronomy (through the MSIP program). The Management Team is advised by an external Advisory Board which meets at least twice a year. 

\textbf{\textit{Partnerships:}} Additional funding for NANOGrav work comes from individual PI grants and multiple prize postdoctoral fellowships, as well as from the Gordon and Betty Moore Foundation (for ultra-wideband receiver development) and Amazon Web Services (to provide high-performance computing for  GW detection efforts). NANOGrav is one of the three primary consortium members of the International Pulsar Timing Array (IPTA), and has partnerships in various ways with the Arecibo Observatory, the GBO, the NRAO, and the CHIME/Pulsar collaboration (Canada).  The IPTA is currently expanding, bringing new partnerships with India, South Africa, and China, and the future will bring additional partnerships with the SKA and its member countries, hopefully as part of a global Radio Alliance, allowing US access to the SKA and other international radio facilities.

\textbf{\textit{Current Status:}}  NANOGrav currently has over 50 Full members and 60 Associate members from $\sim$50 different institutions (undergraduate students not included).  Roughly 60 of those members are graduate students or postdocs, and $\sim$100 undergraduate students are also directly involved in NANOGrav research and observations.  We currently use $\sim$500 hours/year on the GBT, $\sim$750 hours/year on Arecibo, with a smaller test observing program underway with the VLA, to time a total of 77 pulsars (see Figure~\ref{fig:psrs} below), and a new daily-cadence program of our Northern pulsars with CHIME.  The GBT observations are allocated both through the open-skies program and sponsored time funded by the PFC.   An additional several hundred hours per year are used for pulsar surveys critical for growing the array. These include the wide-area GBNCC survey at GBT, the PALFA survey at Arecibo, and radio follow-up of \textit{Fermi} sources.

\begin{figure}[hb]
    \centering
\begin{minipage}[c]{0.6\textwidth}
    \includegraphics[width=\textwidth]{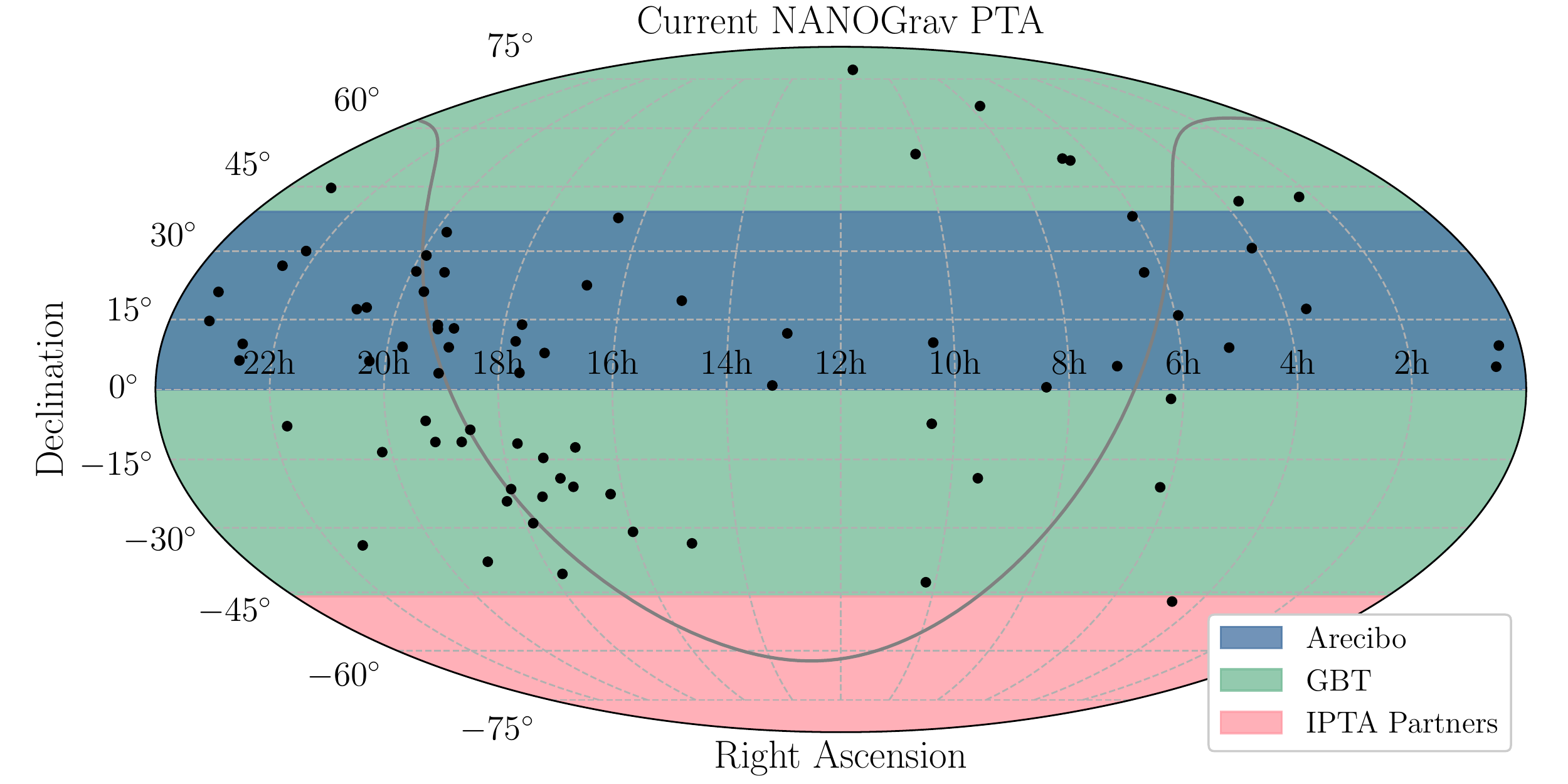}
\end{minipage} 
\hfill
\begin{minipage}[c]{0.38\textwidth}
    \caption{Pulsars currently observed by NANOGrav. PSR~J0437--4715, the southernmost shown, is observed using the VLA only. DSA-2000 will have comparable sky coverage to the two telescopes listed. IPTA partners, such as the Parkes Telescope and MeerKAT, observe pulsars in the Southern sky inaccessible to NANOGrav.
    }
    \label{fig:psrs}
\end{minipage}
\end{figure}

\textbf{\textit{Training:}}  With a very large fraction of our collaboration comprised of junior members, from undergraduates to postdoctoral researchers, scientific education and training are major components of NANOGrav's operations.  Our semi-annual collaboration meetings always feature a student-specific workshop with training in data science as well astrophysics and basic physics, and the annual IPTA meetings, which NANOGrav helps to sponsor, have a student week with similar activities.  Within the collaboration, we have a mentoring program to aid in the career development of our junior members, and  undergraduates are involved in a wide variety of activities, such as observing with the GBT and Arecibo and searching for  MSPs. These activities are both rewarding for the students and critically important components of NANOGrav research.

\textbf{\textit{Outreach:}} NANOGrav's research and training efforts are closely coupled to outreach activities that reach students at the K-12 level and the general public. NANOGrav undergraduate students present SPOT (Space Public Outreach Team) talks at elementary, middle, and high school throughout the US. NANOGrav personnel also participate in the Pulsar Search Collaboratory program, which involves middle and high-school students throughout the state in searches for pulsars using the GBT. NANOGrav also participates in frequent outreach events and members are encouraged to give public lectures. The NANOGrav website also hosts an outreach page with materials for students and the general public. NANOGrav's education and outreach activities are presented in more detail in a State of the Profession White Paper entitled ``NANOGrav Education and Outreach: Growing a Diverse and Inclusive Collaboration for Low-Frequency Gravitational Wave Astronomy''.

\section{Schedule}
\label{sec:schedule}

The timeline for telescope usage and broad science deliverables is shown in Figure~\ref{fig:timeline1}, and will be discussed in more detail in the following two subsections.

\subsection{Schedule for Science Deliverables}

Assuming continued access to the GBT and Arecibo at our requested levels (see next section for more details), our projected high-level science deliverables are:
\begin{enumerate}[topsep=0pt,itemsep=-2pt,partopsep=-12pt]
    \item Detect a stochastic background of nanohertz GWs within the next 3$-$7 years.
    \item Characterize the background in astrophysical terms based on the expected population of SMBBHs within 2$-$3 years of detection. 
    \item Make the first detection of a SMBBH continuous-wave GW system by the end of the decade.
    \item Detect continuous-wave SMBBHs via multi-wavelength follow-up by mid-2030s (and potentially much sooner).
    \item Constrain fundamental physics of cosmology, gravity, and neutron stars, as well as galactic and extra-galactic astrophysics throughout the decade (i.e.~ the synergistic science produced by pulsar timing, in general).
\end{enumerate}

\subsection{Schedule for Data Collection -- Telescope Usage Plan}

The plan for the next decade is divided into two parts. 

\begin{figure}[b!]
    \centering
    \hspace{-2ex}\includegraphics[width=0.51\textwidth]{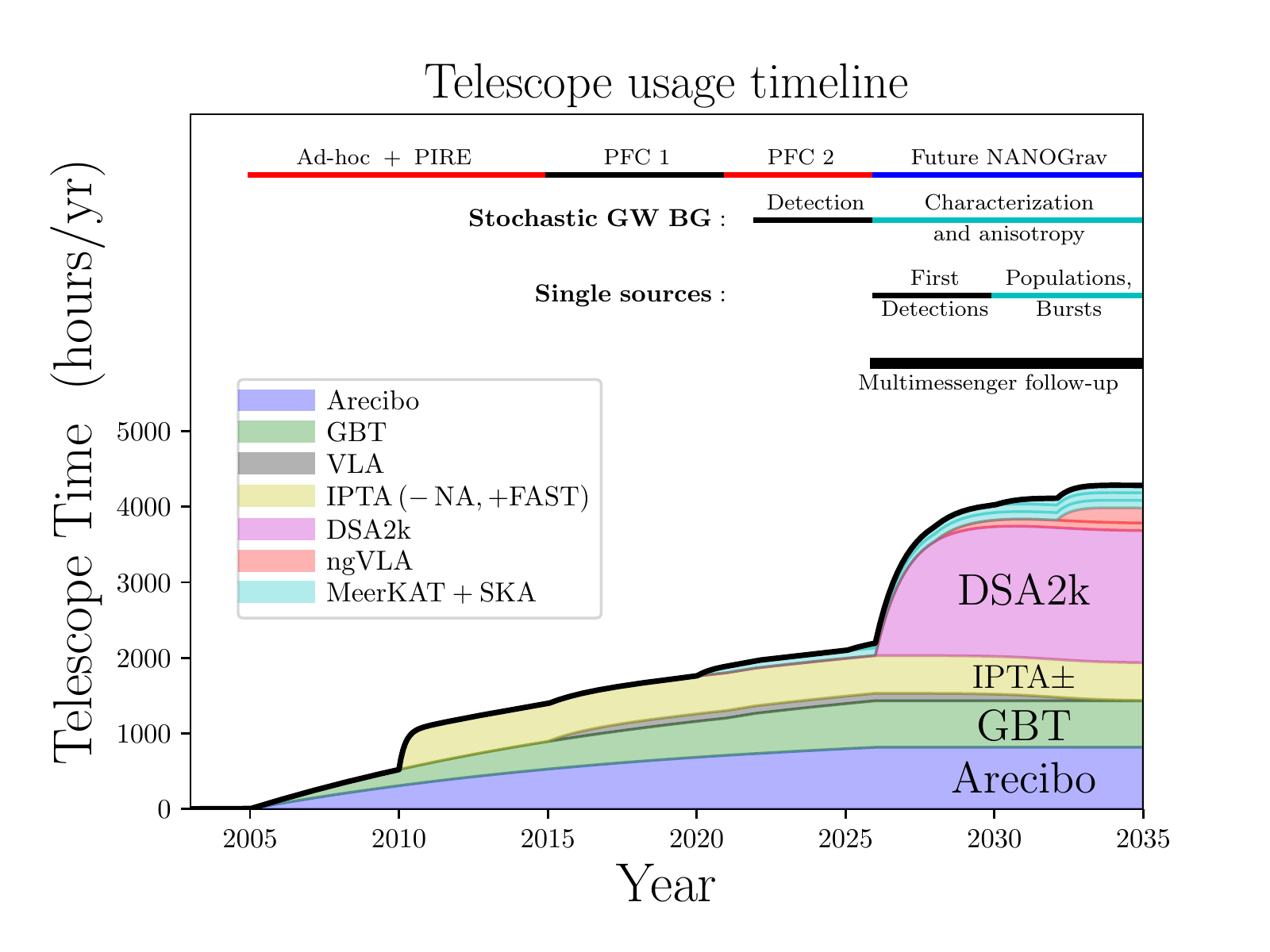}
    \hspace{-2ex}\includegraphics[width=0.51\textwidth]{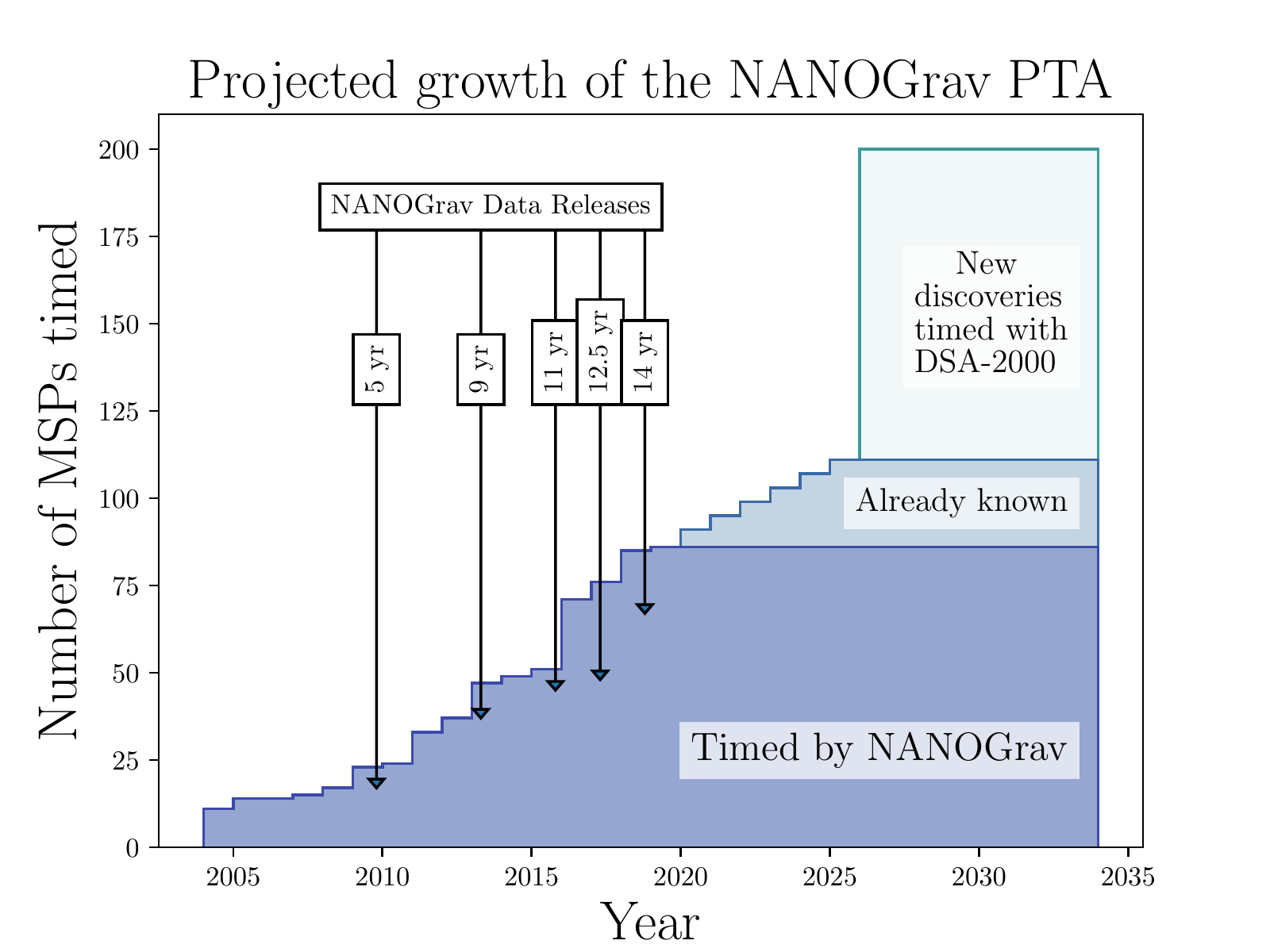}\hspace{0ex}
    \caption{ {\bf Left:} Timeline for NANOGrav's telescope usage for pulsar timing of MSPs.   Additional telescope time is needed in the first half of the decade for search programs to increase the number of MSPs in the pulsar timing array. The time frames for key gravitational-wave science deliverables are also indicated.   Organization and funding of NANOGrav are labeled in the top line, including prior funding under NSF's PIRE program and current funding as a Physics Frontiers Center.  
    {\bf Right:} The past and projected growth of the NANOGrav pulsar timing array. Pulsars with at least 2 years of timing are included in NANOGrav data releases. In future, we predict a ramp up in the number of pulsars observed based on the already-discovered population, constrained by the time available on telescopes at the correct LST ranges. The availability of time on DSA-2000 leads to a large jump in the number of timed pulsars. See discussion in Section~\ref{sec:schedule}.}
    \label{fig:psrbydate}
    \label{fig:timeline1}
\end{figure}

{\bf 2020 -- 2026}:
NANOGrav's observational program comprises a sustained timing program with regular cadence on a growing number of MSPs that results from a coordinated set of pulsar surveys.  During this period we will detect the stochastic background and will transition to characterizing it and detecting individual GW sources. 

In the first half of the decade, MSPs will be timed using the current slate of telescopes (Arecibo, GBT, and the VLA) augmented with CHIME observations for better modeling of interstellar perturbations.  MSPs in the far South will be timed by IPTA partners.  We will also carry out an aggressive campaign to grow the sample of MSPs that we are timing.  The primary MSP search program will use the ALPACA phased-array receiver at Arecibo, now under development under an NSF/MSIP grant (\$5M), with first light expected in 2022.   A large-scale survey of $\sim$5000~deg$^2$, optimized for MSP detection, will extend from 2022--2027 and is expected to discover 50 to 100 MSPs and a wide-variety of other scientifically interesting systems (e.g.~fast radio bursts and relativistic binaries, including possibly a pulsar-black hole system).  Many MSPs will result from continued radio follow-up of {\em Fermi} unassociated sources with the GBT and Arecibo, which have already provided dozens of MSPs for PTA use, as well as from surveys with the LOFAR (Europe) and MeerKAT telescopes whose outcomes will be shared via the IPTA.   Not all MSPs are suitable for GW detection purposes, so pilot observations will be made on each MSP for quality assessment.   The goal is to have a sample that approaches 200 MSPs by roughly the middle the decade.

{\bf 2026 -- 2030}:
NANOGrav's observational program in the latter half of the decade is to time $\sim$200 MSPs in order to better constrain the stochastic GW background and to detect the first individual GW SMBBH sources.

New telescopes, combined with continued GBT and Arecibo observations, will provide the necessary sensitivity and throughput to achieve our goals.  The first is the Deep Synoptic Array (DSA-2000) being proposed to Astro2020 as a multi-purpose telescope with a significant pulsar timing component.   DSA-2000 will provide Arecibo-class sensitivity with 70\% sky coverage and the current agreement is for NANOGrav to use $\sim$1750\,hr/year (25\% of the on-sky time) beginning in $\sim$2026.  The other large facility is the Next Generation VLA (ngVLA), which could have first light with 100\,m-class sensitivity by $\sim$2027--2028.  NANOGrav intends to transition from the current VLA to ngVLA during this time period and grow its usage as ngVLA adds more collecting area -- particularly for faint and/or distant pulsars.  During this period, Phase 1 of the Square Kilometer Array (SKA) should also commence operations, substantially improving the timing precision of Southern pulsars with its Arecibo-class sensitivity.  However, we expect the telescope to be heavily oversubscribed so that high cadence observations will possible on only a relatively small number of objects, using hundreds of hours per year (similar to the ngVLA).

{\bf Beyond 2030}: NANOGrav's goal is sustained timing of NANOGrav and IPTA MSPs to provide very high sensitivity to multiple individual GW sources by 2035.

Ideally GBT and Arecibo observations will continue, and as ngVLA sensitivity increases, usage of the VLA will cease.  The ngVLA is planned to be operating at full sensitivity by 2034, and it will provide higher radio frequency capabilities that will be crucial for good MSPs with large dispersion measures. The DSA-2000 will provide the bulk of NANOGrav's timing (in terms of total hours), and will be supplemented by high sensitivity IPTA telescopes, such as MeerKAT, FAST, and SKA1.  {\bf In the absence of the ngVLA or DSA-2000, continued access to the GBT and Arecibo is critical for U.S.~pulsar and nanohertz GW science.}

\section{Cost Estimates}

The costs of the NANOGrav program over the next decade are outlined in Table~\ref{tab:costtable}. These costs represent the full costs of our program and would almost certainly be funded through multiple sources. Telescope construction costs are detailed in the APC White Paper 
{\it The DSA-2000 - A Radio Survey Camera} (Contacts: G. Hallinan and V. Ravi).  We assume that our personnel needs will be similar to those now, a reasonable assumption given increasingly greater automation in the future. These personnel will likely be supported through a combination of funding sources. The telescope usage profile outlined in Figure~\ref{fig:timeline1} requires roughly 1500 hrs/yr of observations with the GBT and Arecibo for 10 years and roughly 1500 hrs/yr of observations with DSA-2000 for four years, with the remainder relying on other telescopes.  The estimates in Table~\ref{tab:costtable} assume \$2k/hr operating costs for GBT and Arecibo and \$1k/hr operating costs for DSA-2000. We anticipate some, and perhaps even the majority, of these needs to be met through open skies time funded through the NSF AST facilities budget. Our optimal program also requires sophisticated instrumentation, in the form of wideband receivers (in the 0.7--3 GHz range) and backends for Arecibo and DSA-2000, which will likely be funded through a combination of NSF and  private foundations. It is also critical that we continue our education and public outreach activities and provide for both domestic and international meetings. Finally, a sophisticated cyberinfrastructure effort underpins the entire research and education program presented here.

\begin{table}[h]
    \begin{tabular}{l|r}
    \hline
         {\bf Telescope Construction} (for DSA-2000 or equivalent large-number small-diameter array) & \$100M \\
     {\bf  Personnel} (assuming 20 senior personnel, 10 postdocs, 20 graduate students, and 50 undergraduate students)  & \$60M \\
     {\bf Telescope Usage} (assuming usage projected in Figure~\ref{fig:timeline1} and \$2k/hr Arecibo/GBT cost and \$1k/hr DSA-2000 cost)& \$36M \\
     {\bf Instrumentation } (wideband receivers for Arecibo and DSA-2000 and accompanying backends) & 
     \$10M\\
     {\bf Education and Outreach} (SPOT and PSC programs, outreach materials, web resources, student workshops)& \$5M\\
     {\bf Participation Costs }(travel, domestic and international collaboration meetings)& \$5M\\
          {\bf Cyberinfrastructure} (data storage and serving, computational resources) & \$2M\\
     \hline
    {\bf Total} & \$218M\\
    \hline
    \end{tabular}
    \caption{Cost Estimates for the NANOGrav Project for 2020--2030}
    \label{tab:costtable}
\end{table}

\clearpage


\begin{thebibliography}{10}

\bibitem{hd83}
R.~W. {Hellings} and G.~S. {Downs}.
\newblock {Upper limits on the isotopic gravitational radiation background
  frompulsar timing analysis.}
\newblock {\em \apjl}, 265:L39--L42, Feb 1983.

\bibitem{2012MNRAS.427.2780H}
G.~{Hobbs}, W.~{Coles}, R.~N. {Manchester}, M.~J. {Keith}, R.~M. {Shannon},
  D.~{Chen}, M.~{Bailes}, N.~D.~R. {Bhat}, S.~{Burke-Spolaor}, and
  D.~{Champion}.
\newblock {Development of a pulsar-based time-scale}.
\newblock {\em \mnras}, 427(4):2780--2787, Dec 2012.

\bibitem{2010ApJ...720L.201C}
D.~J. {Champion}, G.~B. {Hobbs}, R.~N. {Manchester}, R.~T. {Edwards}, D.~C.
  {Backer}, M.~{Bailes}, N.~D.~R. {Bhat}, S.~{Burke-Spolaor}, W.~{Coles}, and
  P.~B. {Demorest}.
\newblock {Measuring the Mass of Solar System Planets Using Pulsar Timing}.
\newblock {\em \apjl}, 720(2):L201--L205, Sep 2010.

\bibitem{11yrdataset}
Zaven {Arzoumanian}, Adam {Brazier}, Sarah {Burke-Spolaor}, Sydney
  {Chamberlin}, Shami {Chatterjee}, Brian {Christy}, James~M. {Cordes}, Neil~J.
  {Cornish}, Fronefield {Crawford}, H.~{Thankful Cromartie}, et~al.
\newblock {The NANOGrav 11-year Data Set: High-precision Timing of 45
  Millisecond Pulsars}.
\newblock {\em \apjs}, 235(2):37, Apr 2018.

\bibitem{phinney2001}
E.~S. {Phinney}.
\newblock {A Practical Theorem on Gravitational Wave Backgrounds}.
\newblock {\em arXiv e-prints}, pages astro--ph/0108028, Aug 2001.

\bibitem{jaffebacker}
A.~H. {Jaffe} and D.~C. {Backer}.
\newblock {Gravitational Waves Probe the Coalescence Rate of Massive Black Hole
  Binaries}.
\newblock {\em \apj}, 583(2):616--631, Feb 2003.

\bibitem{sesana13}
A.~{Sesana}.
\newblock {Systematic investigation of the expected gravitational wave signal
  from supermassive black hole binaries in the pulsar timing band.}
\newblock {\em \mnras}, 433:L1--L5, Jun 2013.

\bibitem{kh13}
John {Kormendy} and Luis~C. {Ho}.
\newblock {Coevolution (Or Not) of Supermassive Black Holes and Host Galaxies}.
\newblock {\em \araa}, 51(1):511--653, Aug 2013.

\bibitem{mm13}
Nicholas~J. {McConnell} and Chung-Pei {Ma}.
\newblock {Revisiting the Scaling Relations of Black Hole Masses and Host
  Galaxy Properties}.
\newblock {\em \apj}, 764(2):184, Feb 2013.

\bibitem{nano11yr_gwb}
Z.~{Arzoumanian}, P.~T. {Baker}, A.~{Brazier}, S.~{Burke-Spolaor}, S.~J.
  {Chamberlin}, S.~{Chatterjee}, B.~{Christy}, J.~M. {Cordes}, N.~J. {Cornish},
  F.~{Crawford}, et~al.
\newblock {The NANOGrav 11 Year Data Set: Pulsar-timing Constraints on the
  Stochastic Gravitational-wave Background}.
\newblock {\em \apj}, 859(1):47, May 2018.

\bibitem{sej+13}
Xavier {Siemens}, Justin {Ellis}, Fredrick {Jenet}, and Joseph~D. {Romano}.
\newblock {The stochastic background: scaling laws and time to detection for
  pulsar timing arrays}.
\newblock {\em Classical and Quantum Gravity}, 30(22):224015, Nov 2013.

\bibitem{rsg15}
Pablo~A. {Rosado}, Alberto {Sesana}, and Jonathan {Gair}.
\newblock {Expected properties of the first gravitational wave signal detected
  with pulsar timing arrays}.
\newblock {\em \mnras}, 451(3):2417--2433, Aug 2015.

\bibitem{taylor16}
S.~R. {Taylor}, M.~{Vallisneri}, J.~A. {Ellis}, C.~M.~F. {Mingarelli}, T.~J.~W.
  {Lazio}, and R.~{van Haasteren}.
\newblock {Are We There Yet? Time to Detection of Nanohertz Gravitational Waves
  Based on Pulsar-timing Array Limits}.
\newblock {\em \apjl}, 819(1):L6, Mar 2016.

\bibitem{2017MNRAS.471.4508K}
Luke~Zoltan {Kelley}, Laura {Blecha}, Lars {Hernquist}, Alberto {Sesana}, and
  Stephen~R. {Taylor}.
\newblock {The gravitational wave background from massive black hole binaries
  in Illustris: spectral features and time to detection with pulsar timing
  arrays}.
\newblock {\em \mnras}, 471(4):4508--4526, Nov 2017.

\bibitem{SBS2019}
Sarah {Burke-Spolaor}, Stephen~R. {Taylor}, Maria {Charisi}, Timothy {Dolch},
  Jeffrey~S. {Hazboun}, A.~Miguel {Holgado}, Luke~Zoltan {Kelley}, T.~Joseph~W.
  {Lazio}, Dustin~R. {Madison}, Natasha {McMann}, et~al.
\newblock {The astrophysics of nanohertz gravitational waves}.
\newblock {\em \aapr}, 27(1):5, Jun 2019.

\bibitem{Rodriguez2006}
C.~{Rodriguez}, G.~B. {Taylor}, R.~T. {Zavala}, A.~B. {Peck}, L.~K. {Pollack},
  and R.~W. {Romani}.
\newblock {A Compact Supermassive Binary Black Hole System}.
\newblock {\em \apj}, 646(1):49--60, Jul 2006.

\bibitem{2013CQGra..30v4014S}
A.~{Sesana}.
\newblock {Insights into the astrophysics of supermassive black hole binaries
  from pulsar timing observations}.
\newblock {\em Classical and Quantum Gravity}, 30(22):224014, Nov 2013.

\bibitem{2015PhRvD..91h4055S}
Laura {Sampson}, Neil~J. {Cornish}, and Sean~T. {McWilliams}.
\newblock {Constraining the solution to the last parsec problem with pulsar
  timing}.
\newblock {\em \prd}, 91(8):084055, Apr 2015.

\bibitem{2017PhRvL.118r1102T}
Stephen~R. {Taylor}, Joseph {Simon}, and Laura {Sampson}.
\newblock {Constraints on the Dynamical Environments of Supermassive Black-Hole
  Binaries Using Pulsar-Timing Arrays}.
\newblock {\em \prl}, 118(18):181102, May 2017.

\bibitem{2019MNRAS.tmp.1679C}
Siyuan {Chen}, Alberto {Sesana}, and Christopher~J. {Conselice}.
\newblock {Constraining astrophysical observables of Galaxy and Supermassive
  Black Hole Binary Mergers using Pulsar Timing Arrays}.
\newblock {\em \mnras}, page 1679, Jun 2019.

\bibitem{2016ApJ...826...11S}
Joseph {Simon} and Sarah {Burke-Spolaor}.
\newblock {Constraints on Black Hole/Host Galaxy Co-evolution and Binary
  Stalling Using Pulsar Timing Arrays}.
\newblock {\em \apj}, 826(1):11, Jul 2016.

\bibitem{WP_Taylor}
Stephen {Taylor}, Sarah {Burke-Spolaor}, Paul~T. {Baker}, Maria {Charisi},
  Kristina {Islo}, Luke~Z. {Kelley}, Dustin~R. {Madison}, Joseph {Simon}, Sarah
  {Vigeland}, and {Nanograv Collaboration}.
\newblock {Supermassive Black-hole Demographics \&amp;Environments With Pulsar
  Timing Arrays}.
\newblock In {\em \baas}, volume~51, page 336, May 2019.

\bibitem{mls+17}
Chiara M.~F. {Mingarelli}, T.~Joseph~W. {Lazio}, Alberto {Sesana}, Jenny~E.
  {Greene}, Justin~A. {Ellis}, Chung-Pei {Ma}, Steve {Croft}, Sarah
  {Burke-Spolaor}, and Stephen~R. {Taylor}.
\newblock {The local nanohertz gravitational-wave landscape from supermassive
  black hole binaries}.
\newblock {\em Nature Astronomy}, 1:886--892, Nov 2017.

\bibitem{kbh+18}
Luke~Zoltan {Kelley}, Laura {Blecha}, Lars {Hernquist}, Alberto {Sesana}, and
  Stephen~R. {Taylor}.
\newblock {Single sources in the low-frequency gravitational wave sky:
  properties and time to detection by pulsar timing arrays}.
\newblock {\em \mnras}, 477(1):964--976, Jun 2018.

\bibitem{2016ApJ...817...70T}
S.~R. {Taylor}, E.~A. {Huerta}, J.~R. {Gair}, and S.~T. {McWilliams}.
\newblock {Detecting Eccentric Supermassive Black Hole Binaries with Pulsar
  Timing Arrays: Resolvable Source Strategies}.
\newblock {\em \apj}, 817(1):70, Jan 2016.

\bibitem{2019MNRAS.485..248G}
Janna~M. {Goldstein}, Alberto {Sesana}, A.~Miguel {Holgado}, and John {Veitch}.
\newblock {Associating host galaxy candidates to massive black hole binaries
  resolved by pulsar timing arrays}.
\newblock {\em \mnras}, 485(1):248--259, May 2019.

\bibitem{2019MNRAS.485.1579K}
Luke~Zoltan {Kelley}, Zolt{\'a}n {Haiman}, Alberto {Sesana}, and Lars
  {Hernquist}.
\newblock {Massive BH binaries as periodically variable AGN}.
\newblock {\em \mnras}, 485(2):1579--1594, May 2019.

\bibitem{WP_Kelley}
Luke {Kelley}, M.~{Charisi}, S.~{Burke-Spolaor}, J.~{Simon}, L.~{Blecha},
  T.~{Bogdanovic}, M.~{Colpi}, J.~{Comerford}, D.~{D'Orazio}, and M.~{Dotti}.
\newblock {Multi-Messenger Astrophysics With Pulsar Timing Arrays}.
\newblock In {\em \baas}, volume~51, page 490, May 2019.

\bibitem{2019arXiv190611936I}
Kristina {Islo}, Joseph {Simon}, Sarah {Burke-Spolaor}, and Xavier {Siemens}.
\newblock {Prospects for Memory Detection with Low-Frequency Gravitational Wave
  Detectors}.
\newblock {\em arXiv e-prints}, page arXiv:1906.11936, Jun 2019.

\bibitem{2010MNRAS.401.2372V}
Rutger {van Haasteren} and Yuri {Levin}.
\newblock {Gravitational-wave memory and pulsar timing arrays}.
\newblock {\em \mnras}, 401(4):2372--2378, Feb 2010.

\bibitem{2012ApJ...752...54C}
J.~M. {Cordes} and F.~A. {Jenet}.
\newblock {Detecting Gravitational Wave Memory with Pulsar Timing}.
\newblock {\em \apj}, 752:54, June 2012.

\bibitem{2015ApJ...810..150A}
Z.~{Arzoumanian}, A.~{Brazier}, S.~{Burke-Spolaor}, S.~J. {Chamberlin},
  S.~{Chatterjee}, B.~{Christy}, J.~M. {Cordes}, N.~J. {Cornish}, P.~B.
  {Demorest}, and X.~{Deng}.
\newblock {NANOGrav Constraints on Gravitational Wave Bursts with Memory}.
\newblock {\em \apj}, 810(2):150, Sep 2015.

\bibitem{WP_Siemens}
Xavier {Siemens}, Jeffrey {Hazboun}, Paul~T. {Baker}, Sarah {Burke-Spolaor},
  Dustin~R. {Madison}, Chiara {Mingarelli}, Joseph {Simon}, and Tristan
  {Smith}.
\newblock {Physics Beyond the Standard Model With Pulsar Timing Arrays}.
\newblock In {\em \baas}, volume~51, page 437, May 2019.

\bibitem{2008ApJ...685.1304L}
K.~J. {Lee}, F.~A. {Jenet}, and Richard~H. {Price}.
\newblock {Pulsar Timing as a Probe of Non-Einsteinian Polarizations of
  Gravitational Waves}.
\newblock {\em \apj}, 685(2):1304--1319, Oct 2008.

\bibitem{2012PhRvD..85h2001C}
Sydney~J. {Chamberlin} and Xavier {Siemens}.
\newblock {Stochastic backgrounds in alternative theories of gravity: Overlap
  reduction functions for pulsar timing arrays}.
\newblock {\em \prd}, 85(8):082001, Apr 2012.

\bibitem{2015PhRvD..92j2003G}
Jonathan~R. {Gair}, Joseph~D. {Romano}, and Stephen~R. {Taylor}.
\newblock {Mapping gravitational-wave backgrounds of arbitrary polarisation
  using pulsar timing arrays}.
\newblock {\em \prd}, 92(10):102003, Nov 2015.

\bibitem{2018PhRvL.120r1101C}
Neil~J. {Cornish}, Logan {O'Beirne}, Stephen~R. {Taylor}, and Nicol{\'a}s
  {Yunes}.
\newblock {Constraining Alternative Theories of Gravity Using Pulsar Timing
  Arrays}.
\newblock {\em \prl}, 120(18):181101, May 2018.

\bibitem{WP_Fonseca}
Emmanuel {Fonseca}, Paul {Demorest}, Scott {Ransom}, and Ingrid {Stairs}.
\newblock {Fundamental Physics with Radio Millisecond Pulsars}.
\newblock In {\em \baas}, volume~51, page 425, May 2019.

\bibitem{WP_Stinebring}
Dan~R. {Stinebring}, Shami {Chatterjee}, Susan~E. {Clark}, James~M. {Cordes},
  Timothy {Dolch}, Carl {Heiles}, Alex~S. {Hill}, Megan {Jones}, Victoria
  {Kaspi}, and Michael~T. {Lam}.
\newblock {Twelve Decades: Probing the Interstellar Medium from kiloparsec to
  sub-AU scales}.
\newblock In {\em \baas}, volume~51, page 492, May 2019.

\bibitem{2011PhRvD..84d3511B}
Shant {Baghram}, Niayesh {Afshordi}, and Kathryn~M. {Zurek}.
\newblock {Prospects for detecting dark matter halo substructure with pulsar
  timing}.
\newblock {\em \prd}, 84(4):043511, Aug 2011.

\bibitem{2019arXiv190104490D}
Jeff~A. {Dror}, Harikrishnan {Ramani}, Tanner {Trickle}, and Kathryn~M.
  {Zurek}.
\newblock {Pulsar Timing Probes of Primordial Black Holes and Subhalos}.
\newblock {\em arXiv e-prints}, page arXiv:1901.04490, Jan 2019.

\bibitem{2018PhRvD..98j2002P}
Nataliya~K. {Porayko}, Xingjiang {Zhu}, Yuri {Levin}, Lam {Hui}, George
  {Hobbs}, Aleksandra {Grudskaya}, Konstantin {Postnov}, Matthew {Bailes},
  N.~D.~Ramesh {Bhat}, and William {Coles}.
\newblock {Parkes Pulsar Timing Array constraints on ultralight scalar-field
  dark matter}.
\newblock {\em \prd}, 98(10):102002, Nov 2018.

\bibitem{WP_Cordes}
James {Cordes}, Maura~A. {McLaughlin}, and {NANOGrav Collaboration}.
\newblock {Gravitational Waves, Extreme Astrophysics, and Fundamental Physics
  with Precision Pulsar Timing}.
\newblock In {\em \baas}, volume~51, page 447, May 2019.

\bibitem{msmv2013}
C.~M.~F. {Mingarelli}, T.~{Sidery}, I.~{Mandel}, and A.~{Vecchio}.
\newblock {Characterizing gravitational wave stochastic background anisotropy
  with pulsar timing arrays}.
\newblock {\em \prd}, 88(6):062005, Sep 2013.

\bibitem{tmg+2015}
S.~R. {Taylor}, C.~M.~F. {Mingarelli}, J.~R. {Gair}, A.~{Sesana},
  G.~{Theureau}, S.~{Babak}, C.~G. {Bassa}, P.~{Brem}, M.~{Burgay}, and R.~N.
  {Caballero}.
\newblock {Limits on Anisotropy in the Nanohertz Stochastic Gravitational Wave
  Background}.
\newblock {\em \prl}, 115(4):041101, Jul 2015.

\bibitem{WP_Lynch}
Ryan {Lynch}, Paul {Brook}, Shami {Chatterjee}, Timoth {Dolch}, Michael
  {Kramer}, Michael~T. {Lam}, Natalia {Lewand owska}, Maura {McLaughlin}, Nihan
  {Pol}, and Ingrid {Stairs}.
\newblock {The Virtues of Time and Cadence for Pulsars and Fast Transients}.
\newblock In {\em \baas}, volume~51, page 461, May 2019.

\bibitem{blf11}
Brian~J. {Burt}, Andrea~N. {Lommen}, and Lee~S. {Finn}.
\newblock {Optimizing Pulsar Timing Arrays to Maximize Gravitational Wave
  Single-source Detection: A First Cut}.
\newblock {\em \apj}, 730(1):17, Mar 2011.

\bibitem{cal+14}
Brian {Christy}, Ryan {Anella}, Andrea {Lommen}, Lee~Samuel {Finn}, Richard
  {Camuccio}, and Emma {Handzo}.
\newblock {Optimization of NANOGrav's Time Allocation for Maximum Sensitivity
  to Single Sources}.
\newblock {\em \apj}, 794(2):163, Oct 2014.

\bibitem{thk+16}
C.~{Tiburzi}, G.~{Hobbs}, M.~{Kerr}, W.~A. {Coles}, S.~{Dai}, R.~N.
  {Manchester}, A.~{Possenti}, R.~M. {Shannon}, and X.~P. {You}.
\newblock {A study of spatial correlations in pulsar timing array data}.
\newblock {\em \mnras}, 455(4):4339--4350, Feb 2016.

\bibitem{abc+2009}
Melissa {Anholm}, Stefan {Ballmer}, Jolien D.~E. {Creighton}, Larry~R. {Price},
  and Xavier {Siemens}.
\newblock {Optimal strategies for gravitational wave stochastic background
  searches in pulsar timing data}.
\newblock {\em \prd}, 79(8):084030, Apr 2009.

\bibitem{dfg+2013}
P.~B. {Demorest}, R.~D. {Ferdman}, M.~E. {Gonzalez}, D.~{Nice}, S.~{Ransom},
  I.~H. {Stairs}, Z.~{Arzoumanian}, A.~{Brazier}, S.~{Burke-Spolaor}, S.~J.
  {Chamberlin}, et~al.
\newblock {Limits on the Stochastic Gravitational Wave Background from the
  North American Nanohertz Observatory for Gravitational Waves}.
\newblock {\em \apj}, 762(2):94, Jan 2013.

\bibitem{ccs+2015}
Sydney~J. {Chamberlin}, Jolien D.~E. {Creighton}, Xavier {Siemens}, Paul
  {Demorest}, Justin {Ellis}, Larry~R. {Price}, and Joseph~D. {Romano}.
\newblock {Time-domain implementation of the optimal cross-correlation
  statistic for stochastic gravitational-wave background searches in pulsar
  timing data}.
\newblock {\em \prd}, 91(4):044048, Feb 2015.

\bibitem{vit+2018}
Sarah~J. {Vigeland}, Kristina {Islo}, Stephen~R. {Taylor}, and Justin~A.
  {Ellis}.
\newblock {Noise-marginalized optimal statistic: A robust hybrid
  frequentist-Bayesian statistic for the stochastic gravitational-wave
  background in pulsar timing arrays}.
\newblock {\em \prd}, 98(4):044003, Aug 2018.

\bibitem{nano11yr_cw}
K.~{Aggarwal}, Z.~{Arzoumanian}, P.~T. {Baker}, A.~{Brazier}, M.~R. {Brinson},
  P.~R. {Brook}, S.~{Burke-Spolaor}, S.~{Chatterjee}, J.~M. {Cordes}, N.~J.
  {Cornish}, et~al.
\newblock {The NANOGrav 11-Year Data Set: Limits on Gravitational Waves from
  Individual Supermassive Black Hole Binaries}.
\newblock {\em arXiv e-prints}, page arXiv:1812.11585, Dec 2018.

\bibitem{jkc+2018}
Ross~J. {Jennings}, David~L. {Kaplan}, Shami {Chatterjee}, James~M. {Cordes},
  and Adam~T. {Deller}.
\newblock {Binary Pulsar Distances and Velocities from Gaia Data Release 2}.
\newblock {\em \apj}, 864(1):26, Sep 2018.

\bibitem{mab+2018}
Chiara M.~F. {Mingarelli}, Lauren {Anderson}, Megan {Bedell}, and David~N.
  {Spergel}.
\newblock {Improving Binary Millisecond Pulsar Distances with Gaia}.
\newblock {\em arXiv e-prints}, page arXiv:1812.06262, Dec 2018.

\bibitem{dgb+2019}
A.~T. {Deller}, W.~M. {Goss}, W.~F. {Brisken}, S.~{Chatterjee}, J.~M. {Cordes},
  G.~H. {Janssen}, Y.~Y. {Kovalev}, T.~J.~W. {Lazio}, L.~{Petrov}, and B.~W.
  {Stappers}.
\newblock {Microarcsecond VLBI Pulsar Astrometry with PSR{\ensuremath{\pi}} II.
  Parallax Distances for 57 Pulsars}.
\newblock {\em \apj}, 875(2):100, Apr 2019.

\bibitem{tve+16}
S.~R. {Taylor}, M.~{Vallisneri}, J.~A. {Ellis}, C.~M.~F. {Mingarelli}, T.~J.~W.
  {Lazio}, and R.~{van Haasteren}.
\newblock {Are We There Yet? Time to Detection of Nanohertz Gravitational Waves
  Based on Pulsar-timing Array Limits}.
\newblock {\em \apjl}, 819(1):L6, Mar 2016.

\bibitem{2013CQGra..30v4002C}
James~M. {Cordes}.
\newblock {Limits to PTA sensitivity: spin stability and arrival time precision
  of millisecond pulsars}.
\newblock {\em Classical and Quantum Gravity}, 30(22):224002, Nov 2013.

\bibitem{2016ApJ...819..155L}
M.~T. {Lam}, J.~M. {Cordes}, S.~{Chatterjee}, Z.~{Arzoumanian}, K.~{Crowter},
  P.~B. {Demorest}, T.~{Dolch}, J.~A. {Ellis}, R.~D. {Ferdman}, and E.~F.
  {Fonseca}.
\newblock {The NANOGrav Nine-year Data Set: Noise Budget for Pulsar Arrival
  Times on Intraday Timescales}.
\newblock {\em \apj}, 819(2):155, Mar 2016.

\bibitem{2017ApJ...834...35L}
M.~T. {Lam}, J.~M. {Cordes}, S.~{Chatterjee}, Z.~{Arzoumanian}, K.~{Crowter},
  P.~B. {Demorest}, T.~{Dolch}, J.~A. {Ellis}, R.~D. {Ferdman}, and
  E.~{Fonseca}.
\newblock {The NANOGrav Nine-year Data Set: Excess Noise in Millisecond Pulsar
  Arrival Times}.
\newblock {\em \apj}, 834(1):35, Jan 2017.

\bibitem{2016ApJ...821...66L}
M.~T. {Lam}, J.~M. {Cordes}, S.~{Chatterjee}, M.~L. {Jones}, M.~A.
  {McLaughlin}, and J.~W. {Armstrong}.
\newblock {Systematic and Stochastic Variations in Pulsar Dispersion Measures}.
\newblock {\em \apj}, 821(1):66, Apr 2016.

\bibitem{2016MNRAS.457.4421C}
R.~N. {Caballero}, K.~J. {Lee}, L.~{Lentati}, G.~{Desvignes}, D.~J. {Champion},
  J.~P.~W. {Verbiest}, G.~H. {Janssen}, B.~W. {Stappers}, M.~{Kramer}, and
  P.~{Lazarus}.
\newblock {The noise properties of 42 millisecond pulsars from the European
  Pulsar Timing Array and their impact on gravitational-wave searches}.
\newblock {\em \mnras}, 457(4):4421--4440, Apr 2016.

\bibitem{2016MNRAS.458.2161L}
L.~{Lentati}, R.~M. {Shannon}, W.~A. {Coles}, J.~P.~W. {Verbiest}, R.~{van
  Haasteren}, J.~A. {Ellis}, R.~N. {Caballero}, R.~N. {Manchester},
  Z.~{Arzoumanian}, and S.~{Babak}.
\newblock {From spin noise to systematics: stochastic processes in the first
  International Pulsar Timing Array data release}.
\newblock {\em \mnras}, 458(2):2161--2187, May 2016.

\bibitem{dbb+15}
A.~{Dunning}, M.~{Bowen}, M.~{Bourne}, D.~{Hayman}, and S.~L. {Smith}.
\newblock An ultra-wideband dielectrically loaded quad-ridged feed horn for
  radio astronomy.
\newblock In {\em 2015 IEEE-APS Topical Conference on Antennas and Propagation
  in Wireless Communications (APWC)}, pages 787--790, Sep. 2015.

\bibitem{WP_Chang}
Philip {Chang}, Gabrielle {Allen}, Warren {Anderson}, Federica~B. {Bianco},
  Joshua~S. {Bloom}, Patrick~R. {Brady}, Adam {Brazier}, S.~Bradley {Cenko},
  Sean~M. {Couch}, and Tyce {DeYoung}.
\newblock {Cyberinfrastructure Requirements to Enhance Multi-messenger
  Astrophysics}.
\newblock In {\em \baas}, volume~51, page 436, May 2019.

\end{thebibliography}

\end{document}